\documentclass[fleqn]{mn}
\onecolumn
\usepackage{amsfonts}
\usepackage{amsmath}
\usepackage{graphicx}
\usepackage{subfigure}
\usepackage{multicol}

\def\gsim{\;\lower4pt\hbox{${\buildrel\displaystyle >\over\sim}$}\;}
\def\lsim{\;\lower4pt\hbox{${\buildrel\displaystyle <\over\sim}$}\;}
\def\grls{\;\lower4pt\hbox{${\buildrel\displaystyle >\over <}$}\;}

\begin{document}

\title[MHD structures in scale-free discs]
{Structures in a class of magnetized scale-free discs }
\author[Y. Shen, X. Liu, Y.-Q. Lou ]
{Yue Shen$^{1}$, Xin Liu$^{1}$ and Yu-Qing Lou$^{1, 2, 3}$
\\
$^{1}$Physics Department, The Tsinghua Center for Astrophysics
(THCA), Tsinghua University, Beijing 100084, China;\\
$^{2}$National Astronomical Observatories, Chinese Academy
of Sciences, A20, Datun Road, Beijing 100012, China;\\
$^{3}$Department of Astronomy and Astrophysics, The University
of Chicago, 5640 South Ellis Avenue, Chicago, IL 60637 USA. }
\date{Accepted 2004 ... Received 2004 ...;
in original form 2004 ... } \maketitle

\begin{abstract}
We construct analytically stationary global configurations for
both aligned and logarithmic spiral coplanar magnetohydrodynamic
(MHD) perturbations in an axisymmetric background MHD disc with
a power-law surface mass density $\Sigma_0\propto r^{-\alpha}$,
a coplanar azimuthal magnetic field $B_0\propto r^{-\gamma}$, a
consistent self-gravity and
a power-law rotation curve $v_0\propto r^{-\beta}$ where $v_0$ is
the linear azimuthal gas rotation speed. The barotropic equation
of state $\Pi\propto\Sigma^{n}$ is adopted for both MHD background
equilibrium and coplanar MHD perturbations where $\Pi$ is the
vertically integrated pressure and $n$ is the barotropic index.
For a scale-free background MHD equilibrium, a relation exists
among $\alpha$, $\beta$, $\gamma$ and $n$ such that only one
parameter (e.g., $\beta$) is independent. For a linear
axisymmetric stability analysis, we provide global criteria in
various parameter regimes. For nonaxisymmetric aligned and
logarithmic spiral cases, two branches of perturbation modes
(i.e., fast and slow MHD density waves) can be derived once
$\beta$ is specified. To complement the magnetized singular
isothermal disc (MSID) analysis of Lou, we extend the analysis to
a wider range of $-1/4<\beta<1/2$. As an example of illustration,
we discuss specifically the $\beta=1/4$ case when the background
magnetic field is force-free. Angular momentum conservation for
coplanar MHD perturbations and other relevant aspects of our
approach are discussed.
\end{abstract}


\section{ Introduction }

%

Magnetic fields are ubiquitous in diverse astrophysical settings
on various scales ranging from proto-stellar discs, young stellar
objects (YSOs), microquasars, quasars, galaxies and grandiose
astrophysical jets to clusters of galaxies. In some cases, magnetic
fields are relatively weak so that dynamical processes are almost
unaffected by their presence. However, there do exist numerous
cases where magnetic fields are necessary and important for both
dynamics and diagnostics, especially in spiral galaxies and
accretion disc systems (e.g., Sofue et al. 1986; Beck et al. 1996;
Balbus \& Hawley 1998; Tagger \& Pellat 1999; Widrow 2002). In
general, it is challenging to model magnetic fields realistically
because of their complexities (Sofue et al. 1986; Kronberg 1994;
Caunt \& Tagger 2001; Widrow 2002). In the problems involving
protostar formation and disc galaxies, Shu \& Li (1997) presumed
the so-called `isopedic' magnetic field configuration where the
mass-to-magnetic flux ratio in a razor thin disc remains constant.
Using a singular isothermal disc (SID) model (Mestel 1963) that is
isopedically magnetized, Shu et al. (2000) studied stationary
coplanar perturbations and proposed that these global modes are
bifurcations to either secularly or dynamically unstable
configurations. Along a separate yet complementary line, Lou
(2002) studied an azimuthally magnetized singular isothermal disc
(MSID) and derived two different global stationary
magnetohydrodynamic (MHD) perturbation modes. Based on the MSID
model, Lou (2002) explored the manifestation of interlaced optical
and magnetic field spiral arms in the outer portion of a disc with
a nearly flat rotation curve such as the case of the nearby spiral
galaxy NGC 6946 (Beck \& Hoernes 1996; Fan \& Lou 1996; Lou \& Fan
1998a, 2002; Frick et al. 2000, 2001; Lou et al. 2002).

In contexts of spiral galaxies, it is natural and important to consider
a composite disc system consisting of gravitationally coupled gaseous
and stellar discs. This is because various physical processes on
different scales occur in the gaseous disc yet large-scale gas dynamics
and environment are significantly affected by large-scale structures
in the stellar disc (Lou \& Fan 1998b). To construct global coplanar
perturbation structures in a systematic manner, we started with a
composite SID system of two fluid discs without magnetic field (Lou \&
Shen 2003) to generalize the work of Shu et al. (2000) for a single SID.
In terms of the global axisymmetric stability for such a composite SID
system, we provided a more straightforward $D-$criterion (Shen \& Lou
2003) in contrast to the axisymmetric stability criteria of Elmegreen
(1995) and Jog (1996). To be more general than a background SID profile,
we recently constructed global stationary perturbation patterns in a
composite system of two fluid scale-free discs without magnetic field
(Shen \& Lou 2004a, b) to further generalize the work of Syer \&
Tremaine (1996). With proper adaptations, these global stationary
perturbation solutions may be utilized to model large-scale
structures spiral galaxies.

While a scale-free disc has no characteristic scales, one may
impose boundary conditions to describe a modified scale-free disc,
e.g., by cutting a central hole in a disc (e.g., Zang 1976; Evans
\& Read 1998a, b). With such boundary conditions, discrete
eigenfunctions for perturbations may be constructed as growing
normal modes. On the other hand, by specifying a phase relation
for a postulated reflection of spiral waves from the origin
$r=0$, Goodman \& Evans (1999) could also define discrete normal
modes even for an unmodified gaseous SID. Shu et al. (2000)
speculated that the swing amplification process (e.g., Goldreich
\& Lynden-Bell 1965; Julian \& Toomre 1966; Toomre 1977; Fan \&
Lou 1997) across corotation allows a continuum of normal modes
(Lynden-Bell \& Lemos 1993) while proper `boundary conditions'
may select from this continuum a discrete spectrum of unstable
normal modes; they also suggested that zero-frequency (stationary)
disturbances do signal the onset of instability.

On the basis of the work of Shu et al. (2000) for a single
isopedically magnetized SID and of Lou \& Shen (2003) for
a composite SID system without magnetic field, Lou \& Wu
(2004) analyzed coplanar MHD perturbation structures in a
composite SID system with one of the disc being isopedically
magnetized (Shu \& Li 1997). As X-ray emitting hot gases can
freely flow along open magnetic field lines anchored at the
gaseous disc, global stationary solutions constructed by Lou
\& Wu (2004) set
an important preliminary stage for modelling spiral galactic
MHD winds. In particular, if activities of star formation,
starbursts and supernovae etc. are mainly responsible for
producing X-ray emitting hot gases streaming out of the
galactic plane from both sides, we would expect stronger
winds from the circumnuclear starburst region (e.g. Lou et
al. 2001) and along spiral arms than those winds from the
remainder of the galactic disc plane.

Both SID and MSID models offer valuable physical insight for disc
dynamics and belong to a wider family of scale-free discs (Syer \&
Tremaine 1996) where physical variables in the axisymmetric
background equilibrium scale as powers of radius $r$ (e.g., the
rotation curve $\propto r^{-\beta}$). It turns out that coplanar
perturbations in such scale-free discs in a proper range of $\beta$
can be treated globally and analytically (Lemos et al. 1991; Syer
\& Tremaine 1996) without invoking the usual WKBJ or tight-winding
approximation to solve the Poisson equation for density wave
perturbations\footnote{By numerically solving the perturbed
Poisson equation using the Fourier-Bessel transform, R\"udiger \&
Kitchatinov (2000) approached the linear stability of the disc in
terms of an eigenvalue problem and could also treat the density
wave problem globally. Their formulation differs from ours in two
major aspects. Firstly, they include a central mass that mainly
determines the rotation curve (i.e., an approximate Keplerian disc
rotation for a disk mass being much less than the central mass).
In our model, the disc rotation curve is completely controlled
by the disc self-gravity and our approach is semi-analytical.
Secondly, for axisymmetric instabilities, they only revealed ring
fragmentation instabilities when the rotational Mach number becomes
sufficiently high (which is similar to our result). However, our
model analysis further indicates another type of collapse
instabilities (i.e., magneto-rotational Jeans instability) in the
perturbation regime of large radial spatial scales (see Section
3.2.1 for more details).}
(Lin \& Shu 1964; Binney \& Tremaine 1987; Bertin \& Lin 1996). In
other words, it becomes feasible to include azimuthal magnetic fields
in a scale-free disc and to construct global coplanar MHD perturbation
configurations that are stationary in an inertial frame of reference.
Such analytical solutions, still idealized and limited, are making
important further steps and are extremely valuable for bench-marking
numerical codes and for initializing numerical MHD simulations.

For an azimuthally magnetized scale-free disc with an
infinitesimal thickness, the axisymmetric background equilibrium
without radial flow is characterized by following features: a
surface mass density $\Sigma_0\propto r^{-\alpha}$, a rotation
curve $v_0\propto r^{-\beta}$, an azimuthal magnetic field
$B_0\propto r^{-\gamma}$ and a barotropic equation of state
$\Pi=K\Sigma^n$ where $\Pi$ is the vertically integrated
gas pressure with constant $K>0$ for a warm disc and $n>0$ as
the barotropic index. To maintain a radial force balance at
all radii, there exists an explicit relationship among $\alpha$,
$\beta$, $\gamma$ and $n$ such that only one parameter is
independent. As will be seen in Section 2, it is convenient to
use $\beta$ as this independent parameter (Syer \& Tremaine 1996;
Shen \& Lou 2004b).
In order to satisfy relevant constraints, the prescribed range of
$\beta$ turns out to be $-1/4<\beta<1/2$ that includes the
previous special case of $\beta=0$ (i.e., the MSID case) studied
by Lou (2002), Lou \& Zou (2004a) and Lou \& Wu (2004).
In Section 2, we introduce coplanar MHD perturbations and present
the linearized perturbation equations. We construct and analyze
global stationary aligned and logarithmic spiral perturbation
configurations in Section 3 and as an example of illustration, we
discuss specifically the case of $\beta=1/4$ with the background
azimuthal magnetic field being force-free. In Section 4, we derive
phase relationships between perturbation enhancements of surface
mass density and magnetic field for both aligned and logarithmic
spiral cases, and discuss the problem of angular momentum
conservation. Finally, we summarize our results in Section 5.
For the convenience of reference, relevant technical details
are collected in the Appendices.

\section{Formulation of the Problem }

In the magnetofluid approximation, the disc is taken to be
razor-thin (i.e., we use vertically integrated magnetofluid
equations and neglect vertical derivatives of physical variables)
and large-scale stationary aligned and spiral coplanar disturbances
develop in a background MHD rotational equilibrium of axisymmetry
(Syer \& Tremaine 1996; Lou 2002; Lou \& Zou 2004a; Lou \& Wu 2004).
The background magnetic field is taken to be azimuthal to avoid the
magnetic field winding dilemma (Lou \& Fan 1998a). For the sake of
simplicity at this stage, non-ideal effects such as viscosity,
resistivity and thermal diffusion etc. are ignored for large-scale
perturbations.

\subsection{Basic Coplanar MHD Equations in a Cylindrical Disc Geometry}

Using cylindrical coordinates $(r,\theta,z)$, we have the
following two-dimensional nonlinear MHD equations for a
razor-thin disc geometry: the mass conservation equation
\begin{equation}\label{MassConv}
\frac{\partial\Sigma}{\partial t}
+\frac{1}{r}\frac{\partial (r\Sigma u)}
{\partial r}+\frac{1}{r^2}\frac{\partial (\Sigma j)}
{\partial\theta}=0\ ,
\end{equation}
where $\Sigma$ is the surface mass density, $u$ is the radial
bulk flow velocity, $j\equiv rv$ is the specific angular
momentum in the vertical $\hat{z}$-direction and $v$ is the
azimuthal linear velocity; the radial component of the momentum
equation
\begin{equation}\label{RadialMomenEqn}
\begin{split}
\frac{\partial u}{\partial t}+u\frac{\partial u}{\partial r}
+\frac{j}{r^2}\frac{\partial u}{\partial\theta}-\frac{j^2}{r^3}
=-\frac{1}{\Sigma}\frac{\partial\Pi}{\partial r}
-\frac{\partial\phi_{T}}{\partial r}
-\frac{1}{\Sigma}\int\frac{dzB_{\theta}}{4\pi r}
\bigg[\frac{\partial (rB_{\theta})}{\partial r}
-\frac{\partial B_r}{\partial \theta}\bigg]\ ,
\end{split}
\end{equation}
where $\Pi$ is the vertically integrated gas pressure (sometimes
referred to as two-dimensional pressure), $\phi_T$ is the total
gravitational potential including that of an axisymmetric dark
matter halo in contexts of disc galaxies, and $B_{\theta}$ and
$B_r$ are the azimuthal and radial components of the coplanar
magnetic field $\vec B$, respectively; the azimuthal component
of the momentum equation
\begin{equation}\label{AzimuMomenEqn}
\begin{split}
\frac{\partial j}{\partial t}+u\frac{\partial j}{\partial r}
+\frac{j}{r^2}\frac{\partial j}{\partial\theta}
=-\frac{1}{\Sigma}\frac{\partial \Pi}{\partial\theta}
-\frac{\partial \phi_T}{\partial\theta}
+\frac{1}{\Sigma}\int\frac{dzB_r}{4\pi}
\bigg[\frac{\partial (rB_{\theta})}{\partial r}
-\frac{\partial B_r}{\partial\theta}\bigg]\ ;
\end{split}
\end{equation}
the Poisson integral equation
\begin{equation}\label{PoissonInt}
F\phi_T=-G\oint
d\psi\int_0^{\infty}\frac{\Sigma(r^{\prime},\psi,t)
r^{\prime}dr^{\prime}}{[r^{\prime 2}
+r^2-2r^{\prime}r\cos(\psi-\theta)]^{1/2}}\ ,
\end{equation}
where $F\equiv\phi/\phi_T$ is defined as the ratio of the
potential arising from the disc to that arising from the entire
system including a massive dark matter halo that is presumed not
to respond to coplanar perturbations in the disc plane (Syer \&
Tremaine 1996; Shu \& Li 1997; Shu et al. 2000; Lou \& Shen 2003;
Shen \& Lou 2004a, b; Lou \& Zou 2004a) with $F=1$ for a full disc
and $0<F<1$ for partial discs; the divergence-free condition of
magnetic field $\vec B\equiv(B_r,\ B_{\theta},\ 0)$
\begin{equation}\label{DivergFree}
\frac{\partial(rB_r)}{\partial r}+\frac{\partial B_{\theta}}
{\partial\theta}=0\ ;
\end{equation}
the radial component of the magnetic induction equation
\begin{equation}\label{RadiaMInduct}
\frac{\partial B_r}{\partial t}=\frac{1}{r}\frac{\partial}
{\partial\theta}(uB_{\theta}-vB_r)\ ;
\end{equation}
the azimuthal component of the
magnetic field induction equation
\begin{equation}\label{AzimuMInduct}
\frac{\partial B_{\theta}}{\partial t}
=-\frac{\partial }{\partial r}(uB_{\theta}-vB_r)\ ;
\end{equation}
and the barotropic equation of state
\begin{equation}\label{barotropic}
\Pi=K\Sigma^n\ ,
\end{equation}
where constants $K>0$ and $n>0$. Among the three equations
(\ref{DivergFree}), (\ref{RadiaMInduct}) and (\ref{AzimuMInduct}),
we can freely choose two independent ones. By barotropic equation
of state (\ref{barotropic}), the sound speed $a$ is defined by
\begin{equation}\label{soundspeed}
a^2\equiv\frac{d\Pi_0}{d\Sigma_0}=nK\Sigma_0^{n-1}\ ,
\end{equation}
where subscript $0$ indicates the background equilibrium. An
isothermal sound speed corresponds to a barotropic index $n=1$.

\subsection{ Properties of an Axisymmetric Equilibrium }

We now proceed to derive properties of an axisymmetric rotational
MHD equilibrium characterized by physical variables associated with
a subscript $0$. In our notations, such a background equilibrium
has the following power-law scalings: surface mass density of
$\Sigma_0\propto r^{-\alpha}$, a radial velocity profile $u_0=0$ and
$j_0\equiv rv_0\propto r^{1-\beta}$, a purely azimuthal magnetic
field with $B_r^0=0$ and $B_{\theta}^0\equiv B_0\propto r^{-\gamma}$.
Substitutions of these power-law scalings into equations
$(\ref{MassConv})-(\ref{soundspeed})$ lead to the following
condition for the radial force balance, namely
\begin{equation}\label{SFeqn}
v_0^2+\alpha a^2-(1-\gamma)C_A^2=r\frac{d\phi_T^0}{dr}\ ,
\end{equation}
where the Alfv\'en speed $C_A$ in a thin disc is defined by
\begin{equation}\label{Alfven}
C_A^2\equiv\frac{1}{4\pi\Sigma_0}\int B_0^2dz
\end{equation}
and the Poisson integral gives
\begin{equation}\label{phi0total}
F\phi_T^0=-2\pi Gr\Sigma_0{\cal P}_0(\alpha)\ ,
\end{equation}
with the numerical factor ${\cal P}_0(\alpha)$
explicitly defined by
\begin{equation}\label{P0}
{\cal P}_0(\alpha)\equiv\frac{\Gamma(-\alpha/2+1)
\Gamma(\alpha/2 -1/2)}{2\Gamma(-\alpha/2+3/2)\Gamma(\alpha/2)}\
\end{equation}
and $\Gamma(...)$ being the standard gamma function.
This expression (\ref{P0}) can also be included in a
more general form of ${\cal P}_m(\beta)$ as in expression
(\ref{Pm}) derived later with the limiting result of
$2\beta{\cal P}_0(\beta)\rightarrow 1$ as $\beta\rightarrow 0$.

To satisfy the radial force balance equation (\ref{SFeqn}) at
all radii (i.e., scale-free) would necessarily require\footnote{
The magnetic field must obey the scale-free requirement
(\ref{SFcondition}) in general and for the special case
of $\beta=1/4$, we have $\gamma=1$ for the equilibrium
magnetic field to be force-free (e.g., Low \& Lou 1990).
The non-force-free MSID case studied by Lou (2002) has
$\gamma=1/2$, $\beta=0$, $\alpha=1$ and $n=1$. }
\begin{equation}\label{SFcondition}
2\beta=\alpha(n-1)=2\gamma-\alpha=\alpha-1\ ,
\end{equation}
which immediately leads to the explicit expressions of
indices $\alpha$, $\gamma$ and $n$ all in terms of $\beta$
\begin{equation}\label{ParaRelation}
\alpha=1+2\beta\ ,\qquad \gamma=\frac{(1+4\beta )}{2}\ ,
\qquad n=\frac{(1+4\beta)}{(1+2\beta )}\ .
\end{equation}

In this formulation, $\phi_T^0$ remains finite only if $1<\alpha<2$
which implies $0<\beta<1/2$. For the total gravity force arising
from the equilibrium surface mass densities to be finite, a larger
$\beta-$range $\beta\in(-1/2,1/2)$ is allowed. The $\beta-$range is
also constrained by $-1/4<\beta<1/2$ where the first inequality is
required for $n>0$ and the second inequality is imposed such that
the central point mass will not diverge. By these considerations,
the plausible range of $\beta$ falls in $(-1/4,1/2)$ (see Syer \&
Tremaine 1996 and Shen \& Lou 2004a, b).

According to equations (\ref{SFeqn}) and (\ref{ParaRelation}), we
can explicitly write all physical variables of the background
equilibrium state in power-law scalings of $r$ in terms of
parameter $\beta$, namely
\begin{equation}\label{PropEquilib}
\begin{split}
&\Sigma_0=\frac{F A^2[1+D^2-(1-4\beta)q^2/(2+4\beta)]}
{2\pi G(2\beta{\cal P}_0)r^{1+2\beta}}\ ,
\ \ \qquad a^2=\frac{A^2}{(1+2\beta)r^{2\beta}}\ ,
\ \ \qquad C_A^2\equiv q^2a^2=\frac{q^2A^2}{(1+2\beta)r^{2\beta}}\ ,\\
&v_0=\frac{AD}{r^{\beta}}\ ,
\ \ \qquad j_0\equiv rv_0=\frac{AD}{r^{\beta-1}}\ ,
\ \ \qquad \Omega\equiv\frac{j_0}{r^2}=\frac{AD}{r^{1+\beta}}\ ,
\ \ \qquad\kappa\equiv
[(2\Omega/r)d(r^2\Omega)/dr]^{1/2}=[2(1-\beta)]^{1/2}\Omega\ .
\end{split}
\end{equation}
By introducing a reference radius $R$, the constant $A$ is actually
related to a scaled sound speed $a(1+2\beta)^{1/2}R^{\beta}$, the
constant $D=v_0/[a(1+2\beta)^{1/2}]$ is a scaled rotational Mach
number, the constant $q$ is the ratio of the Alfv\'en speed $C_A$
to the sound speed $a$, $\Omega$ and $\kappa$ are the equilibrium
angular speed and the epicyclic frequency, respectively (Lou 2002;
Shen \& Lou 2004a, b). The first expression of equation
(\ref{PropEquilib}) puts an upper limit on the magnetic field
strength, namely, $1+D^2-(1-4\beta)q^2/(2+4\beta)>0$. Alternatively,
once a magnetic field parameter $q^2$ is chosen, the scaled rotational
Mach number $D$ must meet the following physical requirement
\begin{equation}\label{restrt}
D^2>\frac{(1-4\beta)q^2}{(2+4\beta)}-1\
\end{equation}
for a positive equilibrium surface mass density $\Sigma_0$.
Note that for $1/4\le\beta<1/2$, inequality (\ref{restrt}) is
automatically satisfied without actually restricting $q^2$ and
the relevant physical requirement is simply $D^2>0$. Not all
stationary solutions of $D^2$ can satisfy condition (\ref{restrt}).
This then implies limits on allowed magnetic field strength for a
known background rotational MHD equilibrium.

It is also important to note that for the axisymmetric background
rotational MHD equilibrium under consideration, the magnetic force
arising from the azimuthal magnetic field is radially inward when
$-1/4<\beta<1/4$, radially outward when $1/4<\beta<1/2$ and is
zero when $\beta=1/4$ [i.e., the azimuthal magnetic field is
force-free (Lou \& Fan 1998a) and the equilibrium density
distribution reduces to that of the single unmagnetized disc case
(Lemos et al. 1991; Syer \& Tremaine 1996)]. These different
possibilities can be physically understood in terms of the
competition between the magnetic pressure and tension forces.
In order to see this more specifically for the background, we
can write
$-(4\pi\Sigma r)^{-1}\int dzB_{0}\partial (rB_0)/\partial r
=-{\Sigma}^{-1}[(\partial/\partial r)
\int B_0^2dz/(8\pi)+\int B_0^2dz/(4\pi r)]$
referring to equation (\ref{RadialMomenEqn}). The first term
in the square brackets stands for the magnetic pressure force
and the second term stands for the magnetic tension force for
a purely azimuthal magnetic field $B_0(r)$ in the cylindrical
geometry. As $B_0$ scales $r^{-\gamma}$, we have
$(\partial/\partial r)\int B_0^2dz/(8\pi)+\int B_0^2dz/(4\pi r)
=(-\gamma+1)\int B_0^2dz/(4\pi r)$. Therefore for $0<\gamma<1$
or equivalently $-1/4<\beta<1/4$, the magnetic tension force
dominates and the total magnetic force is radially inward; for
$1<\gamma<3/2$ or equivalently $1/4<\beta<1/2$, the magnetic
pressure force dominates and the total magnetic force is
radially outward. The two magnetic forces balance each other
for $\gamma=1$ (or $\beta=1/4$).

\subsection{Coplanar MHD Perturbations in the Disc Plane }

On the basis of the full MHD equations
$(\ref{MassConv})-(\ref{AzimuMInduct})$,
we readily derive the linearized coplanar
MHD perturbation equations as
\begin{equation}\label{Perturb1}
\frac{\partial\Sigma_1}{\partial t}
+\frac{1}{r}\frac{\partial(r\Sigma_0u_1)}{\partial r}
+\Omega\frac{\partial\Sigma_1}{\partial \theta}
+\frac{\Sigma_0}{r^2}\frac{\partial j_1}{\partial\theta}=0\ ,
\end{equation}
\begin{equation}\label{Perturb2}
\begin{split}
\frac{\partial u_1}{\partial t}
+\Omega\frac{\partial u_1}{\partial\theta}
-\frac{2\Omega j_1}{r}=-\frac{\partial}{\partial r}
\bigg(\frac{a^2\Sigma_1}{\Sigma_0}+\phi_1\bigg)
+\frac{(1-4\beta)C_A^2\Sigma_1}
{2\Sigma_0r}-\frac{1}{\Sigma_0}\int\frac{dzB_0}{4\pi
r}\bigg[\frac{\partial(rb_{\theta})}{\partial r}
-\frac{\partial b_r}{\partial\theta}\bigg]
-\frac{1}{\Sigma_0}\int\frac{dzb_{\theta}}{4\pi r}
\frac{\partial(rB_0)}{\partial r}\ ,
\end{split}
\end{equation}
\begin{equation}\label{Perturb3}
\begin{split}
\frac{\partial j_1}{\partial t}
+\frac{r\kappa^2}{2\Omega}u_1+\Omega\frac{\partial j_1}
{\partial\theta}=-\frac{\partial}{\partial
\theta}\bigg(\frac{a^2\Sigma_1}{\Sigma_0}+\phi_1\bigg)
+\frac{1}{\Sigma_0}\int\frac{dzb_r}{4\pi}
\frac{\partial(rB_0)}{\partial r}\ ,
\end{split}
\end{equation}
\begin{equation}\label{Perturb4}
\phi_1=-G\oint
d\psi\int_0^{\infty}\frac{\Sigma_1(r^{\prime},\psi,t)
\ r^{\prime}dr^{\prime}}{[r^{\prime 2}+r^2
-2r^{\prime}r\cos(\psi-\theta)]^{1/2}},
\end{equation}
\begin{equation}\label{Perturb5}
\frac{\partial(rb_r)}{\partial r}
+\frac{\partial b_{\theta}}{\partial\theta}=0\ ,
\end{equation}
\begin{equation}\label{Perturb6}
\frac{\partial b_r}{\partial t}
=\frac{1}{r}\frac{\partial}{\partial\theta}(u_1B_0-r\Omega b_r)\ ,
\end{equation}
\begin{equation}\label{Perturb7}
\frac{\partial b_{\theta}}{\partial t}
=-\frac{\partial}{\partial r}(u_1B_0-r\Omega b_r)\ ,
\end{equation}
where we use subscript $1$ to denote associations
with small disturbances in physical variables and
$\vec b\equiv (b_r,b_{\theta},0)$ stands for the
coplanar magnetic field perturbation. As the background
rotational MHD equilibrium is stationary and axisymmetric,
these perturbed physical variables can be decomposed in
terms of Fourier harmonics with the periodic dependence
$\exp(i\omega t-im\theta)$ where $\omega$ is the angular
frequency and $m$ is an integer for azimuthal variations.
More specifically, we write
\begin{equation}\label{Fourier}
\begin{split}
&\Sigma_1=S(r)\exp(\hbox{i}\omega t-\hbox{i}m\theta)\ ,\quad\qquad
u_1=U(r)\exp(\hbox{i}\omega t-\hbox{i}m\theta) ,\quad\qquad
j_1=J(r)\exp(\hbox{i}\omega t-\hbox{i}m\theta)\ ,\\
&\phi_1=V(r)\exp(\hbox{i}\omega t-\hbox{i}m\theta)\ ,\quad\qquad
b_r=R(r)\exp(\hbox{i}\omega t-\hbox{i}m\theta)\ ,\quad\qquad
b_{\theta}=Z(r)\exp(\hbox{i}\omega t-\hbox{i}m\theta)\ ,
\end{split}
\end{equation}
where $S(r)$, $U(r)$, $J(r)$, $V(r)$, $R(r)$ and $Z(r)$ are
all radial variations of the corresponding perturbed physical
variables and can be complex in general for possible radial
oscillations. Without loss of generality, we take $m\ge 0$
in our analysis.

With Fourier decomposition (\ref{Fourier}), it is
then straightforward to cast coplanar MHD perturbation
equations $(\ref{Perturb1})-(\ref{Perturb7})$ into the
following forms of
\begin{equation}\label{Perturb01}
\hbox{i}(\omega-m\Omega)S+\frac{1}{r}\frac{\partial}{\partial r}
(r\Sigma_0U)-\frac{\hbox{i}m\Sigma_0}{r^2}J=0\ ,
\end{equation}
\begin{equation}\label{Perturb02}
\begin{split}
\hbox{i}(\omega-m\Omega)U-\frac{2\Omega
J}{r}=-\frac{\partial\Phi}{\partial r}
+\frac{(1-4\beta)C_A^2S}{2\Sigma_0r}
-\frac{1}{\Sigma_0}\int\frac{dzZ}{4\pi r}
\frac{\partial(rB_0)}{\partial r}
-\frac{1}{\Sigma_0}\int\frac{dzB_0}{4\pi r}
\bigg[\frac{\partial(rZ)}{\partial r}+\hbox{i}mR\bigg]\ ,
\end{split}
\end{equation}
where $\Phi\equiv a^2S/\Sigma_0+V$ is a short-hand notation, and
\begin{equation}\label{Perturb03}
\hbox{i}(\omega-m\Omega)J+\frac{r\kappa^2}{2\Omega}U=\hbox{i}m\Phi+
\frac{1}{\Sigma_0}\int\frac{dzR}{4\pi}\frac{\partial(rB_0)}
{\partial r}\ ,
\end{equation}
\begin{equation}\label{Perturb04}
V(r)=-G\oint
d\psi\int_0^{\infty}\frac{S(r^{\prime})\cos(m\psi)r^{\prime}dr^{\prime}}
{(r^{\prime 2}+r^2-2rr^{\prime}\cos\psi)^{1/2}}\ ,
\end{equation}
\begin{equation}\label{Perturb05}
\frac{\partial(rR)}{\partial r}-\hbox{i}mZ=0\ ,
\end{equation}
\begin{equation}\label{Perturb06}
\hbox{i}(\omega-m\Omega)R+\frac{\hbox{i}mB_0}{r}U=0\ ,
\end{equation}
\begin{equation}\label{Perturb07}
\hbox{i}\omega Z=\frac{\partial}{\partial r}(r\Omega R)
-\frac{\partial}{\partial r}(B_0U)\ ,
\end{equation}
where equation (\ref{Perturb07}) can be derived by combining
equations (\ref{Perturb05}) and (\ref{Perturb06}).

We now rearrange the time-dependent coplanar MHD perturbation
equations by taking $m\ge 1$ from equations
(\ref{Perturb01})$-$(\ref{Perturb07}); and the special case of
$m=0$ will be analyzed in details at the end of this subsection.

From equations (\ref{Perturb05}) and
(\ref{Perturb06}), we readily obtain
\begin{equation}\label{ZR}
\begin{split}
Z=-\frac{\hbox{i}}{m}\frac{\partial(rR)}{\partial r}\ ,
\qquad\qquad R=-\frac{mB_0U}{r(\omega-m\Omega)}\ .
\end{split}
\end{equation}
A substitution of equation (\ref{ZR}) into the radial
and azimuthal components of the momentum equation
(\ref{Perturb02}) and (\ref{Perturb03}) yields
\begin{equation}\label{radial}
\begin{split}
\hbox{i}(\omega-m\Omega)U-\frac{2\Omega J}{r}
=-\frac{\partial\Phi}{\partial r}
+\frac{(1-4\beta)C_A^2S}{2\Sigma_0r}
-\hbox{i}C_A^2\bigg[\frac{\partial^2}
{\partial r^2}+\frac{1-12\beta}{2r}\frac{\partial}{\partial r}
+\frac{2\beta(1+4\beta)-m^2}{r^2}\bigg]
\bigg(\frac{U}{\omega-m\Omega}\bigg)\ ,
\end{split}
\end{equation}
and
\begin{equation}\label{azimuthal}
\hbox{i}(\omega-m\Omega)J+\frac{r\kappa^2}{2\Omega}U
=\hbox{i}m\Phi-\frac{mC_A^2(1-4\beta)}{2r}
\bigg(\frac{U}{\omega-m\Omega}\bigg)\ ,
\end{equation}
respectively.

In order to construct global solutions without the WKBJ
approximation, we are mainly interested in stationary
configurations with zero pattern speed $\omega/m=0$, also
referred to as neutral $\omega=0$ modes (e.g., Syer \&
Tremaine 1996; Shu et al. 2000; Lou 2002; Shen \& Lou 2004a, b;
Lou \& Zou 2004a). By setting $\omega=0$ and taking $m\ge 1$,
equations (\ref{Perturb01}), (\ref{ZR}), (\ref{radial}) and
(\ref{azimuthal}) can be readily reduced to
\begin{equation}\label{stationary1}
m\Omega S+\frac{1}{r}\frac{\partial}{\partial r}
(r\Sigma_0\hbox{i}U)+\frac{m\Sigma_0}{r^2}J=0\ ,
\end{equation}
\begin{equation}\label{stationary2}
Z=-\frac{\partial}{\partial r}
\bigg(\frac{B_0\hbox{i}U}{m\Omega}\bigg)\ ,
\qquad\qquad\qquad
R=\frac{B_0U}{r\Omega}\ ,
\end{equation}
\begin{equation}\label{stationary3}
\begin{split}
m\Omega \hbox{i}U+\frac{2\Omega J}{r}=\frac{\partial\Phi}
{\partial r} -\frac{(1-4\beta)C_A^2S}{2\Sigma_0r}
-\frac{C_A^2}{m}\bigg[\frac{\partial^2}{\partial r^2}
+\frac{1-12\beta}{2r}\frac{\partial}{\partial r}
+\frac{2\beta(1+4\beta)-m^2}{r^2}\bigg]\frac{\hbox{i}U}{\Omega}\ ,
\end{split}
\end{equation}
\begin{equation}\label{stationary4}
m\Omega J+\frac{r\kappa^2}{2\Omega}\hbox{i}U=-m\Phi
+\frac{C_A^2(1-4\beta)}{2r}\frac{\hbox{i}U}{\Omega}\ .
\end{equation}

Equations (\ref{stationary1}) through (\ref{stationary4}) together
with equation (\ref{Perturb04}) are the basic MHD perturbation
equations for constructing stationary non-axisymmetric $m\ge 1$
configurations of both aligned and unaligned logarithmic spiral
cases. Note that aligned and spiral stationary global solutions
both involve propagations of fast and slow MHD density waves
(Lou 2002).

We next present the basic coplanar MHD perturbation equations
for stationary axisymmetric $m=0$ configurations with or without
radial propagations. Starting from equations (\ref{Perturb01}),
(\ref{Perturb06}), (\ref{Perturb07}), (\ref{radial}) and
(\ref{azimuthal}) by setting $m=0$ and $\omega\neq 0$, we obtain
\begin{equation}\label{m0}
\begin{split}
&\omega S=\frac{1}{r}\frac{\partial}{\partial r}
(r\Sigma_0\hbox{i}U)\ ,\\
&\omega^2 \hbox{i}U-\frac{2\omega\Omega J}{r}
=-\omega\frac{\partial\Phi}{\partial r}
+\omega\frac{(1-4\beta)C_A^2S}{2\Sigma_0r}
-C_A^2\bigg[\frac{\partial^2}{\partial r^2}
+\frac{1-12\beta}{2r}\frac{\partial}{\partial r}
+\frac{2\beta(1+4\beta)}{r^2}\bigg]\hbox{i}U\ ,\\
&\omega J=\frac{r\kappa^2}{2\Omega}\hbox{i}U\ ,\\
&R=0\ ,\qquad\qquad \omega Z=\frac{\partial}{\partial
r}(B_0\hbox{i}U)\ .
\end{split}
\end{equation}
This set of equations is applicable as we proceed to investigate
stationary axisymmetric coplanar MHD perturbations (Lou \& Zou
2004a, b). For coplanar hydrodynamic
perturbations or coplanar MHD perturbations in an isopedically
magnetized SID (Shu et al. 2000), there is no essential difference
for changing the order of limiting procedure for $m=0$ and
$\omega\rightarrow 0$. However, for MSID with a coplanar azimuthal
background magnetic field, the results would be different by
changing the order of limiting procedure for $m=0$ and
$\omega\rightarrow 0$ (Lou 2002; Lou \& Fan 2002; Lou \& Zou
2004a).

\section{ Stationary Aligned and Logarithmic Spiral Configurations }

To solve the Poisson equation connecting the perturbed surface mass
density and the perturbed gravitational potential, we may consistently
assume disturbances to carry proper scale-free forms (Lynden-Bell
\& Lemos 1993) for aligned and logarithmic spiral perturbations. For
aligned perturbations, $S(r)$ contains only an amplitude variation
in $r$ of perturbed surface mass density and does not involve phase
variation in $r$ so that the maximum density perturbations at
different radii line up in the azimuth. For spiral perturbations in
comparison, in addition to an amplitude variation in $r$, $S(r)$ also
involves a phase variation in $r$ such that a spiral pattern emerges.

\subsection{Aligned Global Coplanar MHD Configurations}

For aligned global coplanar MHD perturbations, we select
those perturbations that carry the same power-law variation
in $r$ of the background equilibrium, namely
\begin{equation}\label{alignedPert}
S=\sigma r^{-1-2\beta}\ ,
\qquad\qquad
V=-2\pi GrS{\cal P}_m(\beta)\ ,
\end{equation}
where $\sigma$ is a small constant coefficient and the
numerical factor ${\cal P}_m(\beta)$ is given explicitly by
\begin{equation}\label{Pm}
{\cal P}_m(\beta)=
\frac{\Gamma(m/2-\beta+1/2)\Gamma(m/2+\beta)}
{2\Gamma(m/2-\beta+1)\Gamma(m/2+\beta+1/2)}\ ,
\end{equation}
where $-m/2<\beta<(m+1)/2$ (Qian 1992; Syer \& Tremaine 1996; Shen
\& Lou 2004a, b). In the limit of $\beta\rightarrow 0$, we simply
have ${\cal P}_m=1/m$ consistent with earlier results on SIDs and
MSIDs (Shu et al. 2000; Lou 2002; Lou \& Shen 2003; Lou \& Zou
2004a; Lou \& Wu 2004). In fact, a more general class of
self-consistent potential-density pairs satisfying the Poisson
integral (\ref{Perturb04}) is also available; see footnote 3 in
Shen \& Lou (2004b).

\subsubsection{Aligned $m=0$ Case }

For aligned neutral modes of axisymmetry ($m=0$), we can start from
equation (\ref{m0}) by setting $\omega=0$; this implies $U=R=0$ and
gives no constraints on $Z$. As $S\propto r^{-1-2\beta}$, the
scale-free condition requires $Z\propto r^{-\gamma}$ with
$\gamma=1/2+2\beta$ and $J\propto r^{1-\beta}$. It turns out that
such stationary coplanar MHD perturbations are simply alternative
states for the axisymmetric background equilibrium with a proper
rescaling factor (Shu et al. 2000; Lou 2002). As this is somewhat
trivial, we now turn our attention mainly to cases of
non-axisymmetric $m\ge 1$ stationary aligned coplanar
MHD perturbations.

\subsubsection{Aligned $m\ge 1$ Cases }

For the potential-density pair
$S\propto r^{-1-2\beta}$ and $V=-2\pi G{\cal P}_mrS$, we have
$U\propto r^{-\beta}$ and $J\propto r^{1-\beta}$ according to
equations $(\ref{stationary1})-(\ref{stationary4})$. It then
follows that
\begin{equation}\label{alignedmge1}
\begin{split}
&m\Omega S-\frac{3\beta}{r}
\Sigma_0\hbox{i}U+\frac{m\Sigma_0}{r^2}J=0\ ,\\
&m\Omega \hbox{i}U+\frac{2\Omega}{r}J=\frac{2\beta a^2}{r\Sigma_0}
\bigg(\frac{r\Sigma_0}{a^2}2\pi G{\cal P}_m-1\bigg)S
-\frac{(1-4\beta)C_A^2}{2\Sigma_0r}S
-\frac{C_A^2[(1-4\beta)^2-2m^2]}{2m\Omega r^2}\hbox{i}U\ ,\\
&m\Omega J+r(1-\beta)\Omega
\hbox{i}U=\frac{ma^2}{r\Sigma_0}\bigg(\frac{r\Sigma_0}{a^2}2\pi G
{\cal P}_m-1\bigg)rS+\frac{C_A^2(1-4\beta)}{2\Omega r}\hbox{i}U\ ,\\
&Z=-\bigg(\frac{1}{2}-2\beta\bigg)\frac{B_0\hbox{i}U}{m\Omega r}\ ,
\qquad\qquad\qquad R=\frac{B_0U}{\Omega r}\ .
\end{split}
\end{equation}
Rearranging the first three equations of (\ref{alignedmge1}),
we obtain a set of three algebraic equations, namely
\begin{equation}\label{SUJ}
\begin{split}
a_1S+b_1\hbox{i}U+c_1J=0\ ,
\qquad\qquad a_2S+b_2\hbox{i}U+c_2J=0\ ,
\qquad\qquad a_3S+b_3\hbox{i}U+c_3J=0\ ,
\end{split}
\end{equation}
where coefficients $a_i$, $b_i$, $c_i$
($i=1,2,3$) are explicitly defined by
\begin{equation}\label{abc}
\begin{split}
&a_1\equiv m\Omega\ ,
\qquad\qquad
b_1\equiv -\frac{3\beta\Sigma_0}{r}\ ,
\qquad\qquad
c_1\equiv \frac{m\Sigma_0}{r^2}\ ,\\
&a_2\equiv\frac{(1-4\beta)C_A^2}{2\Sigma_0r}-2\beta\bigg[2\pi G
{\cal P}_m(\beta)-\frac{a^2}{\Sigma_0 r}\bigg]\ ,
\quad\qquad
b_2\equiv m\Omega+\frac{[(1-4\beta)^2-2m^2]C_A^2}{2m\Omega r^2}\ ,
\quad\qquad
c_2\equiv\frac{2\Omega}{r}\ ,\\
&a_3\equiv -mr\bigg[2\pi G{\cal P}_m(\beta)
-\frac{a^2}{\Sigma_0 r}\bigg]\ ,
\qquad\qquad
b_3\equiv r(1-\beta)\Omega-\frac{(1-4\beta)C_A^2}{2\Omega r}\ ,
\qquad\qquad
c_3\equiv m\Omega\ .
\end{split}
\end{equation}
For nontrivial solutions of $\{S\ ,\ \hbox{i}U\ ,\ J\}$,
the determinant of algebraic equations (\ref{SUJ}) must
vanish, namely
\begin{equation}\label{aligned0}
a_1-\frac{b_1(a_2c_3-a_3c_2)}{(b_2c_3-b_3c_2)}
+\frac{c_1(a_2b_3-a_3b_2)}{(b_2c_3-b_3c_2)}=0\ .
\end{equation}
Using coefficient definitions (\ref{abc}) in condition
(\ref{aligned0}), we obtain a more informative form of
stationary dispersion relation
\begin{equation}\label{aligned1}
\begin{split}
\frac{a^2-2\pi G{\cal P}_m\Sigma_0r-\Omega^2r^2}
{(1/2-2\beta)C_A^2-(1+2\beta)\Omega^2r^2}=&\frac{\Omega^2r^2
[6\beta a^2+(2-2\beta)\Omega^2r^2-12\beta\pi G{\cal P}_m
\Sigma_0r-(1-4\beta)C_A^2]}{[m^2\Omega^2r^2
-C_A^2(m^2-1/2-8\beta^2+4\beta)]
[(1+2\beta)\Omega^2r^2-C_A^2(1/2-2\beta)]}\
\\
&\qquad\qquad\qquad
-\frac{4\beta a^2-8\beta\pi G{\cal P}_m\Sigma_0r
+(1-4\beta)C_A^2}{2[m^2\Omega^2r^2
-C_A^2(m^2-1/2-8\beta^2+4\beta)]}\ .
\end{split}
\end{equation}
Note that equation (\ref{aligned1}) takes a simpler form in the
limit of $\beta\rightarrow 0$ (i.e., a flat rotation curve of a
SID) and reduces to equation (3.1.9) of Lou (2002) as expected.

By using background MHD equilibrium variables (\ref{PropEquilib})
in condition (\ref{aligned1}), we obtain the stationary dispersion
relation for global aligned $m\ge 1$ coplanar MHD configurations
in the form of a quadratic equation in terms of $y\equiv D^2$,
namely
\begin{equation}\label{aligned}
C_2y^2+C_1y+C_0=0\ ,
\end{equation}
where the three coefficients $C_i$
($i=1,2,3$) are defined by
\begin{equation}\label{alignC2C1C0}
\begin{split}
&C_2\equiv {\cal H}_m\ ,\\
&C_1\equiv -\bigg[\frac{(3-4\beta)m^2
-(1-4\beta)(1-6\beta+4\beta^2)}{2(1+2\beta)}
F{\cal C}{\cal P}_m+m^2
-(1-4\beta)(2-\beta)\bigg]q^2
-{\cal A}_m(1-F {\cal C}{\cal P}_m)\ ,\\
&C_0\equiv -\frac{1-4\beta}{4(1+2\beta)}
\bigg[\frac{-2m^2+(1-4\beta)(1-2\beta)}{1+2\beta}
F{\cal C}{\cal P}_m+1-4\beta\bigg]q^4
-\frac{-2m^2+(1-4\beta)(1-2\beta)}{2(1+2\beta)}
(1-F{\cal C}{\cal P}_m)q^2\ ,
\end{split}
\end{equation}
with auxiliary parameters explicitly defined by
\begin{equation}\label{ABCD}
\begin{split}
{\cal A}_m\equiv m^2+4\beta-4\beta^2\ ,\qquad
{\cal B}_m\equiv (1+2\beta)(m^2-2+2\beta)\ ,\qquad
{\cal C}\equiv\frac{1+2\beta}{2\beta{\cal P}_0}\ ,\qquad
{\cal H}_m\equiv  F{\cal C}{\cal P}_m{\cal A}_m+{\cal B}_m\ \\
\end{split}
\end{equation}
which are exactly the same as those defined in Shen \& Lou (2004b)
for aligned cases of global perturbations in a purely hydrodynamic
composite disc system. Here, the effect of coplanar magnetic field
is associated with the $q^2$ parameter in definitions
(\ref{alignC2C1C0}). The determinant of quadratic equation
(\ref{aligned}), $\Delta\equiv C_1^2-4C_2C_0$, depends on various
parameters and can become negative under some circumstances (i.e.,
no physical solutions for $y\equiv D^2$). For $\Delta\ge 0$, the
two real solutions of $y\equiv D^2$ to equation (\ref{aligned}) are
\begin{equation}\label{Solalign}
y_1=\frac{-C_1+\Delta^{1/2}}{2C_2}\
\qquad\hbox{ and }\qquad
y_2=\frac{-C_1-\Delta^{1/2}}{2C_2}\ .
\end{equation}

As a first check of necessary consistency, we consider the
limiting case of vanishing magnetic field, that is, $q=0$.
Then, equation (\ref{aligned}) simply reduces to
\begin{equation}\label{alignedNoM}
{\cal H}_my^2-{\cal A}_m(1-F{\cal C}{\cal P}_m)y=0\ ,
\end{equation}
which gives a non-trivial solution
\begin{equation}\label{alignedNoMSol}
D^2=y_1=\frac{(1- F{\cal C}
{\cal P}_m){\cal A}_m}{{\cal H}_m}\ .
\end{equation}
This $D^2$ solution (\ref{alignedNoMSol}) is exactly
the same result for the case of a single scale-free
gas disc without magnetic field (see Syer \& Tremaine
1996 and subsection 3.1 of Shen \& Lou 2004b).

As a second check of necessary consistency, we consider another
limiting case of $\beta\rightarrow 0$ which corresponds to the
MSID case with a flat rotation curve studied by Lou (2002). One
can readily show that equation (\ref{aligned}) reduces to equation
(3.2.11) of Lou (2002) as expected.

In contrast to the case of a single fluid disc studied by Syer \&
Tremaine (1996) without azimuthal magnetic fields, there are now
two branches of $D^2$ solutions of algebraic equation (\ref{aligned})
in general. A $D^2$ solution is considered to be physical when it
satisfies both the condition $D^2>0$ and inequality (\ref{restrt}).
When the magnetic field becomes sufficiently weak, among these two
branches of $D^2$ solutions, $y_1$ is the counterpart of the single
disc case and $y_2$ is additional due to the coupling between the
surface mass density and magnetic field.
We recall that in a composite system of one stellar disc and one
gaseous disc coupled through the mutual gravitational interaction,
there are also two branches of $D^2$ solutions (Lou \& Fan 1998b,
2000; Lou \& Shen 2003; Shen \& Lou 2004b). In a composite
stellar-gaseous disc system, the two different classes of
perturbation modes correspond to either in-phase or out-of-phase
of surface mass density perturbation enhancements (see also
Lou \& Wu 2004).
In a single azimuthally magnetized disc, the two coplanar MHD
modes $y_1$ and $y_2$ will be distinguished by either in-phase
or out-of-phase perturbation enhancements of the surface mass
density and the azimuthal magnetic field in the WKBJ or
tight-winding approximation. In the WKBJ regime, the $y_1$ and
$y_2$ branches correspond to stationary fast and slow MHD
density waves, respectively (Fan \& Lou 1996; Lou \& Fan 1998a).
Although the phase relationships for perturbation enhancements
of the surface mass density and the azimuthal magnetic field
become $\pm\pi/2$ for open spiral structures (Lou 2002; Lou \&
Fan 2002), we shall still refer to $y_1$ and $y_2$ solutions as
global stationary fast and slow MHD density waves, respectively.
We shall come back to this discussion of `phase relationship'
in more details in Section 4.

\subsubsection{Force-Free Magnetized Discs with $\beta=1/4$ }

As an example of illustration, we now focus on the specific case
of $\beta=1/4$ with inequality (\ref{restrt}) automatically
satisfied. In such a magnetized disc system, the rotation curve
scales as $r^{-1/4}$, the surface mass density scales as $r^{-3/2}$,
the barotropic index is $n=4/3$ and the azimuthal magnetic field
scales as $r^{-1}$ (i.e., the background equilibrium magnetic field
is force-free). For this specific case, the determinant $\Delta$ of
algebraic equation (\ref{aligned}) remains always positive for both
full and partial discs (see proofs in Appendix A). As seen from $D^2$
solutions (\ref{Solalign}), the sign of $C_2$ determines whether
$y_1$/$y_2$ is the upper/lower branch or reverse. For $m\ge 2$, we
always have $C_2>0$ and therefore $y_1$ and $y_2$ remain to be the
higher and lower branches, respectively. The situation becomes
somewhat involved for the $m=1$ case when the sign of $C_2$ is
dependent on the value of $ F$ parameter. We show in Fig. 1 the
$y\equiv D^2$ solution branches for $m=1$ and $m=2$ cases,
respectively, versus the gravitational potential ratio $ F$.

For the $m=1$ case, there exists a diverging point at
$ F_c=3/[7{\cal C}{\cal P}_1(1/4)]$ for $y_1\equiv D^2$
solution branch where $C_2={\cal H}_1=0$. The value of this critical
$ F_c$ is numerically determined to be $ F_c\simeq 0.6842$.
More specifically, we have $C_2<0$ for $0<F< F_c$, while we have
$C_2>0$ for $F>F_c$. This sign variation of $C_2$ is indicated
by the relative locations of $y_1$ and $y_2$ solution branches as
shown in panel $(a)$ of Fig. 1. In other words, for $0<F<F_c$,
$y_1$ remains negative and the only physical solution is $y_2$. On
the other hand for $F>F_c$, both $y_1$ and $y_2$ solution branches
become physically plausible.

For $\beta=1/4$, the explicit form of algebraic
equation (\ref{aligned}) becomes
\begin{equation}\label{align0d25}
\bigg[{F}{\cal C}{\cal P}_m
\bigg(m^2+\frac{3}{4}\bigg)
+\frac{3}{2}\bigg(m^2-\frac{3}{2}\bigg)\bigg]D^4
-\bigg[\bigg(\frac{2{F}{\cal C}{\cal P}_m}{3}
+1\bigg)m^2q^2+\bigg(m^2+\frac{3}{4}\bigg)(1-{F}
{\cal C}{\cal P}_m)\bigg]D^2+\frac{2}{3}(1-{F}
{\cal C}{\cal P}_m)m^2q^2=0\ ,
\end{equation}
where auxiliary parameters ${\cal P}_m$, ${\cal A}_m$,
${\cal B}_m$, ${\cal C}$ and ${\cal H}_m$ are all evaluated
at $\beta=1/4$. It is fairly straightforward to show that
$0<{\cal C}{\cal P}_m<1$ for all $m\ge 1$ when $\beta=1/4$
(Shen \& Lou 2004b). As $0<F\le 1$, we then have
$0<1-{F}{\cal C}{\cal P}_m<1$.

\begin{figure}
\centering \subfigure[$m=1$, $\beta=1/4$, $q=0,\ 1$]{
\includegraphics[scale=0.42]{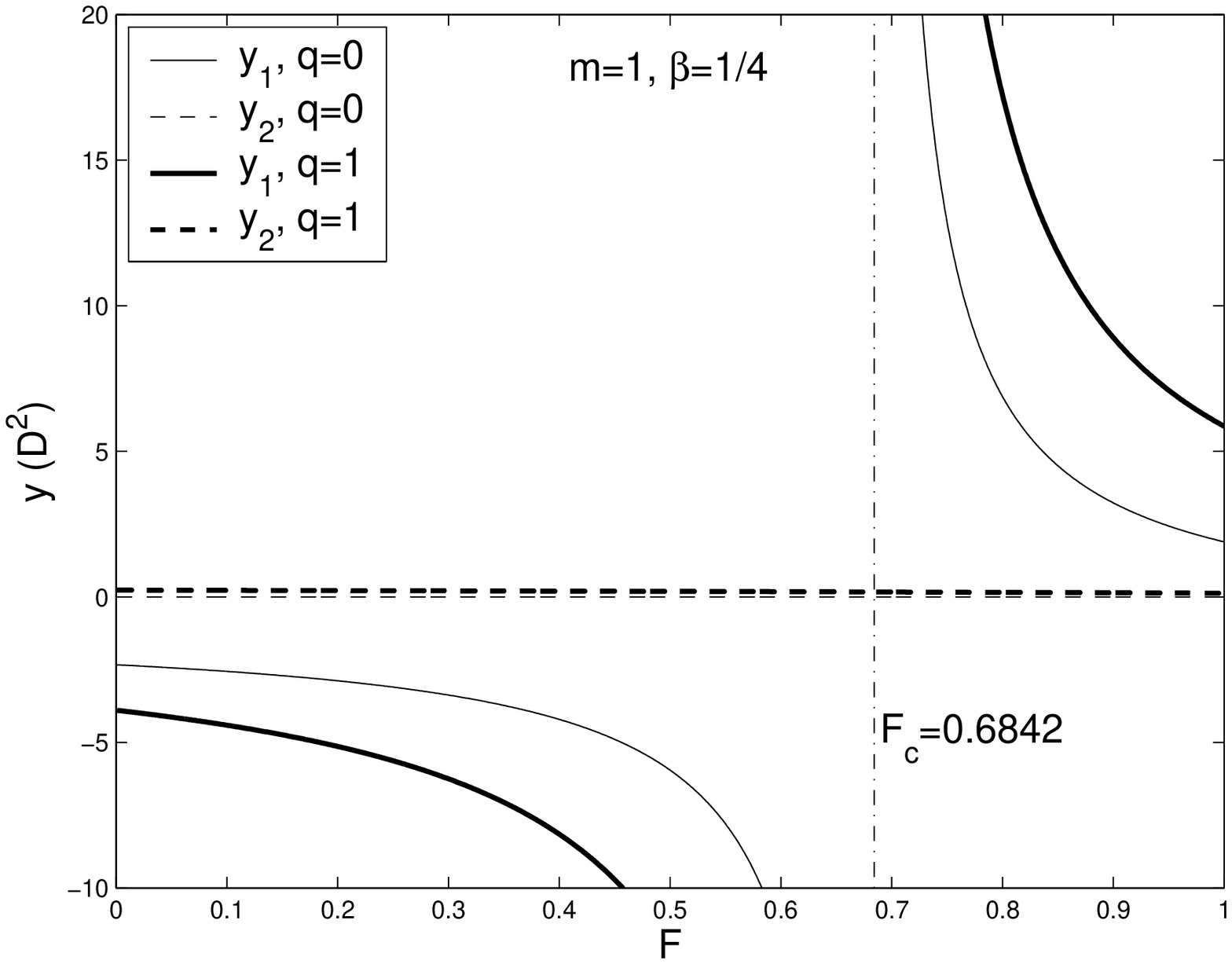}}
\subfigure[$m=2$, $\beta=1/4$, $q=0,\ 1$]{
\includegraphics[scale=0.42]{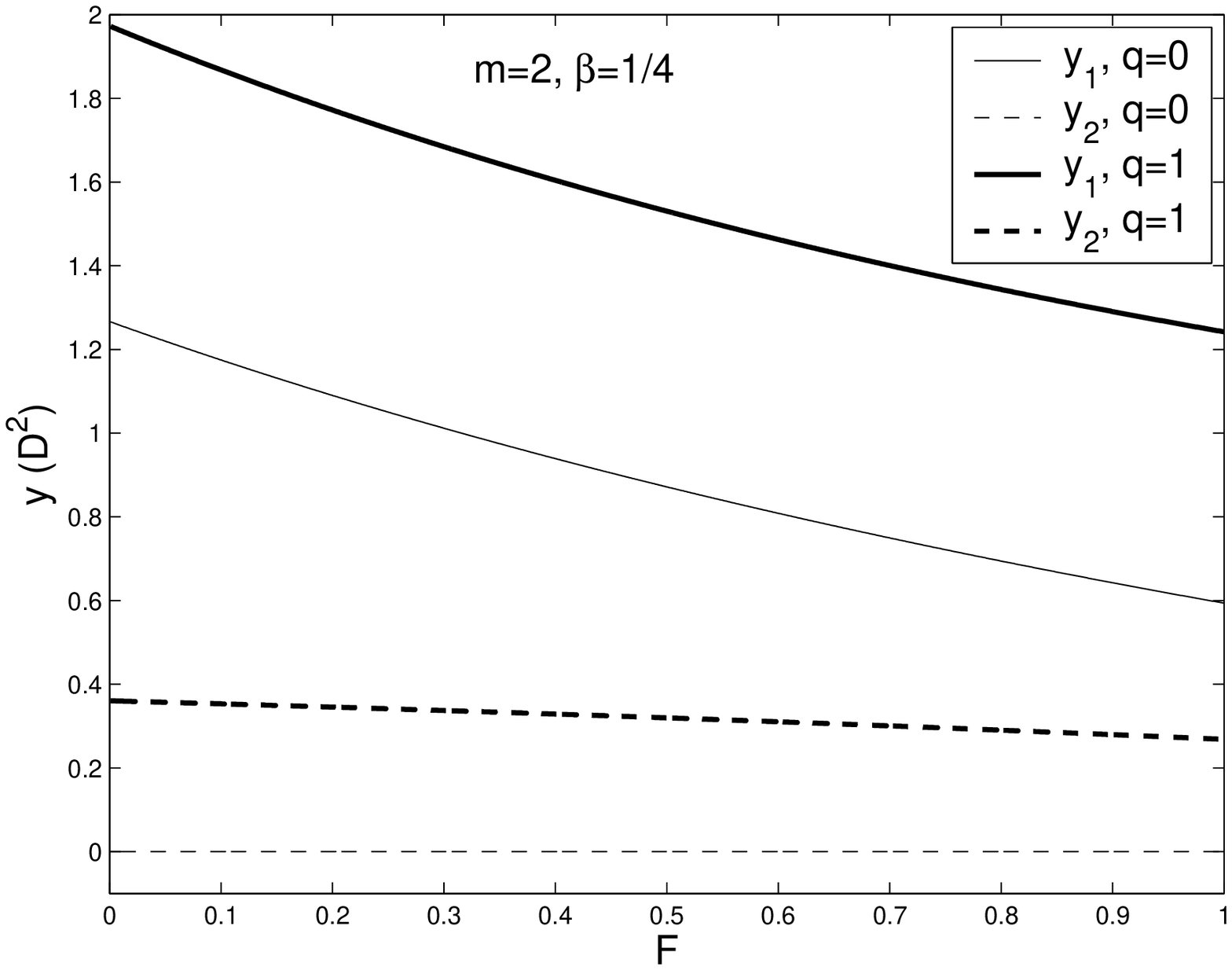}}
\caption{Two solution branches for $y\equiv D^2$ of algebraic
equation (\ref{aligned}) for $m=1,\ 2$, $\beta=1/4$ and $q=0,\ 1$.
The $y_1$ branch is plotted in solid lines and the $y_2$ branch in
dash lines. {\it Panel (a)}: For the $m=1$ case, the relative
locations of the two branches switch across the diverging point at
${F}_c=0.6842$. Physical solutions are those above the $y=0$ axis
marked by the light dash line. For $0<{F}<{F}_c$, only the $y_2$
branch (the dash lines) is physical. For physical solutions, the
enhancement of magnetic field will raise both branches but with
different rate. The $y_1$ branch (the solid lines) is apparently
influenced by the presence of the coplanar magnetic field, while
the $y_2$ branch (the dash lines) is slightly shifted.
{\it Panel (b)}: For the $m=2$ case, $y_1$ and $y_2$ solutions
remain always to be the upper and lower branches, respectively,
and both branches are physical in the range of $0<{F}<1$. Both
branches are raised by the enhancement of the coplanar magnetic
field strength. The $m>2$ cases are qualitatively similar to
the $m=2$ case shown in panel (b) here. }
\end{figure}

\subsection{ Logarithmic Spiral Configurations
for Global Coplanar MHD Perturbations }

In our formulation, stationary surface mass density perturbations
are characterized by $S(r)\exp(-\hbox{i}m\theta)$. For aligned
coplanar MHD perturbations, we took $S\propto r^{-\varepsilon}$
where $\varepsilon$ is a real exponent. In Section 3.1, we have
chosen $\varepsilon=\alpha=1+2\beta$ that carries the same
power-law dependence as the equilibrium disc does. For a complex
$\varepsilon$, a perturbation pattern would appear spiral, that
is, a logarithmic spiral in the form of $S\propto
r^{-\Re(\varepsilon)} \exp [-\hbox{i}\Im(\varepsilon)\ln r]$ where
$\Re(\varepsilon)$ and $\Im(\varepsilon)$ are real and imaginary
parts of $\varepsilon$. To ensure the gravitational potential
arising from this perturbed surface mass density being finite,
we should require $-m+1<\Re(\varepsilon)<m+2$ (e.g., Qian 1992).
In fact, there exists a more general class of self-consistent
potential-density pairs as indicated in footnote 7 of Shen \&
Lou (2004b). However, for the analysis of axisymmetric stability
problem in Section 3.2.1, equation (\ref{spirPerturb}) is used
such that dispersion relation (\ref{m0U}) derived later on is
real on both sides (e.g., Lemos et al. 1991; Syer \& Tremaine
1996; Shu et al. 2000).
To be specific here, we consistently take the logarithmic spiral
surface mass density perturbations and the resulting gravitational
potentials as (Kalnajs 1971; Lemos et al. 1991; Shu et al. 2000;
Lou 2002; Lou \& Shen 2003; Shen \& Lou 2004b; Lou \& Zou 2004a)
\begin{equation}\label{spirPerturb}
\begin{split}
S=\sigma r^{-3/2}\exp(\hbox{i}\xi\ln r) \
\qquad\qquad\hbox{ and }\qquad\qquad\
V=-2\pi GrS{\cal N}_m(\xi)\ ,
\end{split}
\end{equation}
where $\sigma$ is a small constant, $\xi$ is a
parameter related to the radial wavenumber and the
Kalnajs function ${\cal N}_m(\xi)$ is defined by
\begin{equation}
{\cal N}_m(\xi)=\frac{1}{2}
\frac{\Gamma(m/2+\hbox{i}\xi/2+1/4)\Gamma(m/2-\hbox{i}\xi/2+1/4)}
{\Gamma(m/2+\hbox{i}\xi/2+3/4)\Gamma(m/2-\hbox{i}\xi/2+3/4)}
\end{equation}
(Kalnajs 1971). As ${\cal N}_m(\xi)$ is an even function of $\xi$,
a consideration of $\xi\ge 0$ would suffice. In our convention of
notations, $\xi>0$ and $\xi<0$ correspond to leading and trailing
logarithmic spiral waves, respectively, and $\xi\equiv kr$ relates
the parameter $\xi$ to the radial wavenumber $k$. We note that
${\cal N}_m(\xi)$ decreases monotonically with increasing $\xi>0$,
and $0<{\cal N}_m(\xi)<1$ for $m\ge 1$; for $m=0$, ${\cal N}_0(\xi)$
is positive and can be greater than 1 when $\xi$ becomes sufficiently
small.

The choice of unaligned perturbations is not unique, and the
perturbation potential-density pairs may involve $\beta$
parameter such that the background plus coplanar MHD
perturbations are altogether scale-free (e.g., Syer \&
Tremaine 1996). As an example of illustration,
we here presume coplanar logarithmic spiral perturbations
defined by the density-potential pair (\ref{spirPerturb}) in
the following analysis. For the specific $\beta=1/4$ case, the
background surface mass density and the perturbed surface mass
density bear the same radial scaling $\propto r^{-3/2}$.

\subsubsection{ Marginal Axisymmetric Stability Curves }

We start from the axisymmetric coplanar MHD perturbation equation
(\ref{m0}) with $\omega\neq 0$. For radial oscillations with a
logarithmic potential-density pair, we presume a surface mass
density $S\propto r^{-3/2+\hbox{i}\xi}$, a gravitational potential
$V=-2\pi G{\cal N}_0rS$ and thus a radial speed $U\propto r^{1/2+
\hbox{i}\xi+2\beta}$ according to the first of equation (\ref{m0}).
After proper rearrangements, the set of equations (\ref{m0}) can be
combined to form a single equation in terms of $U(r)$, namely
\begin{equation}\label{m0U}
\begin{split}
(\omega^2-\kappa^2)U&=\bigg[\bigg(\frac{a^2}{r^2}
-\frac{2\pi G{\cal N}_0\Sigma_0}{r}\bigg)
\bigg(\frac{1}{4}+\xi^2\bigg)+\frac{C_A^2}{r^2}
\bigg(\frac{1}{4}+\xi^2-\beta\bigg)\bigg]U\\
&=\frac{a^2}{r^2}\bigg\{\bigg[1-{F}{\cal C}
{\cal N}_0\bigg(1+D^2-\frac{1-4\beta}{2+4\beta}q^2\bigg)\bigg]
\bigg(\frac{1}{4}+\xi^2\bigg)
+q^2\bigg(\frac{1}{4}+\xi^2-\beta\bigg)\bigg\}U\ ,
\end{split}
\end{equation}
where ${\cal C}=(1+2\beta)/(2\beta{\cal P}_0)$,
$a^2/r^2=\kappa^2/[(1+2\beta)(2-2\beta)D^2]$ and
$\omega^2$ is extremely small.

By inspection, equation (\ref{m0U}) bears similar features of
the dispersion relation in the classic WKBJ regime involving
magnetic field. As the right-hand side of equation (\ref{m0U})
is real, axisymmetric instabilities first set via neutral
$\omega=0$ modes. It should be emphasized that the scale-free
condition is met only for neutral modes. With $\omega^2=0$,
the marginal stability curves are then given by
\begin{equation}\label{MSC}
D^2=\frac{\{1-{F}{\cal C}{\cal N}_0[1-(1-4\beta)q^2/
(2+4\beta)]\}(\xi^2+1/4)+q^2(\xi^2-\beta+1/4)}{{F}
{\cal C}{\cal N}_0(\xi^2+1/4)+(1+2\beta)(-2+2\beta)}\ .
\end{equation}
As a check of necessary consistency, the axisymmetric
marginal stability curve for an MSID when $\beta=0$
(${\cal C}=1$) then gives
\begin{equation}\label{MSCbeta0}
D^2=\frac{[1+q^2-{F}{\cal N}_0(1-q^2/2)]
(\xi^2+1/4)}{{F}{\cal N}_0(\xi^2+1/4)-2}\ .
\end{equation}
One can readily show\footnote{We recall that inequality (\ref{restrt})
is automatically satisfied for $1/4\le\beta<1/2$ and we need only to
examine the range of $-1/4<\beta<1/4$. When the denominator of the
right-hand side of equation (\ref{MSC}) is negative, the numerator
must also be negative for $D^2>0$. It then follows that
$1-(1-4\beta)q^2/(2+4\beta)>0$ and inequality (\ref{restrt}) is met.
When the denominator of the right-hand side of equation (\ref{MSC})
is positive, it is easy to show that inequality (\ref{restrt}) holds.
As solution (\ref{MSCbeta0}) is a special case of solution (\ref{MSC}),
inequality (\ref{restrt}) is thus met. }
that stationary dispersion relations (\ref{MSC}) and (\ref{MSCbeta0})
automatically satisfy inequality (\ref{restrt}) for physical solutions
$D^2>0$. In the MSID case with different values of $q$ parameter for
magnetic field strengths, one readily obtains the marginal stability
curves that separate two distinct unstable regimes, namely, the
collapse regime for large radial spatial scales and the ring
fragmentation regime for relatively short radial wavelengths (Lemos
et al. 1991; Shen \& Lou 2003, 2004a). There exists a critical
$\xi=\xi_c$ where $D^2$ diverges and this $\xi_c$ is determined
by $F{\cal N}_0(\xi_c^2+1/4)=2$. For $\xi<\xi_c$, we have
$F{\cal N}_0(\xi^2+1/4)-2<0$ and thus an enhancement of $q$
tends to suppress the collapse boundary. For $\xi>\xi_c$, we have
$F{\cal N}_0(\xi^2+1/4)-2>0$ and thus an enhancement of $q$ tends
to raise the ring fragmentation boundary. In short, an enhancement
of ring magnetic field tends to reduce dangers of both collapse
and ring fragmentation instabilities in the MSID case.

While the enhancement of ring magnetic field tends to suppress
ring fragmentation instabilities in general, it only suppresses
collapse instabilities for $-1/4<\beta<1/4$ and tends to aggravate
collapse instabilities for $1/4<\beta<1/2$ (see Fig. 2 for
details).
It has been shown that the magnetic force is radially inward and
outward for $-1/4<\beta<1/4$ and $1/4<\beta<1/2$, respectively.
Why does the trend of variations for the collapse-stability appear
seemingly paradoxical? This situation can be understood because the
unperturbed background is in an MHD radial force balance and the
surface mass density, the rotation speed and the magnetic field
are coupled explicitly through condition (\ref{PropEquilib}).
Furthermore, for fast MHD density waves with $m=0$ (Lou \& Fan
1998a), the gas pressure and magnetic pressure together is
associated with the radial wavenumber squared, while
the surface mass density and self-gravity are associated linearly
with the radial wavenumber (see the first line of dispersion
relation \ref{m0U} with small $\omega^2$). For $-1/4<\beta<1/4$,
the surface mass density becomes smaller for stronger magnetic field
strength, while for $1/4<\beta<1/2$, the situation reverses. For the
ring fragmentation instability at relatively large radial wavenumbers,
the dominant magnetic pressure effect tends to stabilize the disc. For
the Jeans collapse instability at relatively small radial wavenumbers,
the dominant self-gravity effect is proportional to the background
surface mass density; the roles of magnetic field for two different
situations can then be readily understood. It is the coupling effect
between the surface mass density and the magnetic field of the
background that gives rise to this collapse feature.

The marginal stability curves for scale-free discs with
$\beta\in(-1/4,1/2)$ and without magnetic fields are
\begin{equation}\label{MSCq0}
D^2=\frac{(1-{F}{\cal C}{\cal N}_0)(\xi^2+1/4)}{{F}
{\cal C}{\cal N}_0(\xi^2+1/4)+2(2\beta+1)(\beta-1)}\ ,
\end{equation}
that are consistent with those for the single-disc case
of Shen \& Lou (2004b; e.g., see their subsection 3.2.1).

The marginal axisymmetric stability curves generally consist two
branches, namely, the collapse branch and the ring-fragmentation
branch (Lemos et al. 1991; Shu et al. 2000; Lou 2002; Lou \& Shen
2003; Shen \& Lou 2003). The lowest value of $D^2$ for stability
is the maximum of the collapse branch and the highest value of
$D^2$ for stability is the minimum of the ring-fragmentation
branch. We note that the maximum of the collapse branch is always
located at $\xi=0$ for $\beta\ge -0.104$, while for $\beta<-0.104$
the maximum of the collapse branch may locate at $\xi>0$ with
$\xi=0$ corresponding to a local minimum of $D^2$. It was proven in
Appendix C of Shen \& Lou (2004b) that $\xi=0$ is always a local
extremum for $D^2$. Variations of the stable range for $D^2$ with
parameters $F$, $\beta$ and $q$ are shown in panels (a) and (b) of
Fig. 2. Finally we note that the axisymmetric destabilization
involves only the fast MHD wave branch because the slow MHD wave
branch is negative and thus unphysical. For non-axisymmetric
($m\neq 0$) perturbations, both fast and slow MHD density waves
are possible and stationary global configurations may represent
transitions from stable to unstable situations (e.g., Shu et al.
2000). Here, the stability of our slow MHD density waves differs
from those models involving finite disc thickness where either
vertical or horizontal weak magnetic fields may induce
Velikhov-Chandrasekhar-Balbus-Hawley instability and
magneto-rotational instabilities (MRI) through slow MHD modes
(Chandrasekhar 1961; Balbus \& Hawley 1998; Kim \& Ostriker
2000; Lou, Yuan \& Fan 2001).

It would be interesting to recall the axisymmetric stability
analysis in the WKBJ approximation. The local dispersion relation for
fast MHD waves in an azimuthal magnetic field is (Fan \& Lou 1996)
\begin{equation}\label{DRWKBJ}
(\omega-m\Omega)^2=\kappa^2+k^2(C_A^2+a^2-2\pi G\Sigma_0/|k|)\ ,
\end{equation}
where $k$ is the radial wavenumber and other notations
have the conventional meanings. Equation (\ref{DRWKBJ})
gives the marginal stability curve for axisymmetric
$m=0$ perturbations as
\begin{equation}\label{stability}
\kappa^2+k^2(C_A^2+a^2-2\pi G\Sigma_0/|k|)=0\ .
\end{equation}
By using scale-free disc equilibrium condition
(\ref{PropEquilib}) and inserting $\xi\cong |k|r$ for large
$k$ and $\xi$ into dispersion relation (\ref{stability}),
we readily obtain the corresponding marginal stability
curve in the WKBJ regime as
\begin{equation}
D^2=\frac{(1+q^2)\xi^2-F{\cal C}
[1-(1-4\beta)q^2/(2+4\beta)]\xi}
{F{\cal C}\xi+(1+2\beta)(-2+2\beta)}\ ,
\end{equation}
which bears a strong resemblance to expression (\ref{MSC}) of
our global analysis, especially in view of the asymptotic
expression for ${\cal N}_m(\xi)$, namely
\begin{equation}\label{asymN}
{\cal N}_m(\xi)\approx(m^2+\xi^2+1/4)^{-1/2}
\end{equation}
when $m^2+\xi^2\gg 1$. Another equivalent way of this comparison
is to note the parallel terms in the WKBJ condition
(\ref{stability}) and in the first line of the corresponding
global condition (\ref{m0U}), also in reference to the above
asymptotic expression for ${\cal N}_m(\xi)$. A specific
illustrating example of such a comparison is shown in Fig. 3 with
parameters $\beta=0$ (${\cal C}=1$; flat rotation curve), $F=1$
(full disc) and $q=0.1$. In the short-wavelength limit (i.e.,
$\xi\gg 1$), the WKBJ approximation is well justified and the
results are qualitatively consistent, while in the long-wavelength
of Jeans collapse regime, the WKBJ analysis differs from the
global exact treatment significantly.

\begin{figure}
\centering \subfigure[Stability boundary curves
of $\beta$ versus $D^2$ for full discs of $F=1$
with different values of $q$ parameter.]{
\includegraphics[scale=0.42]{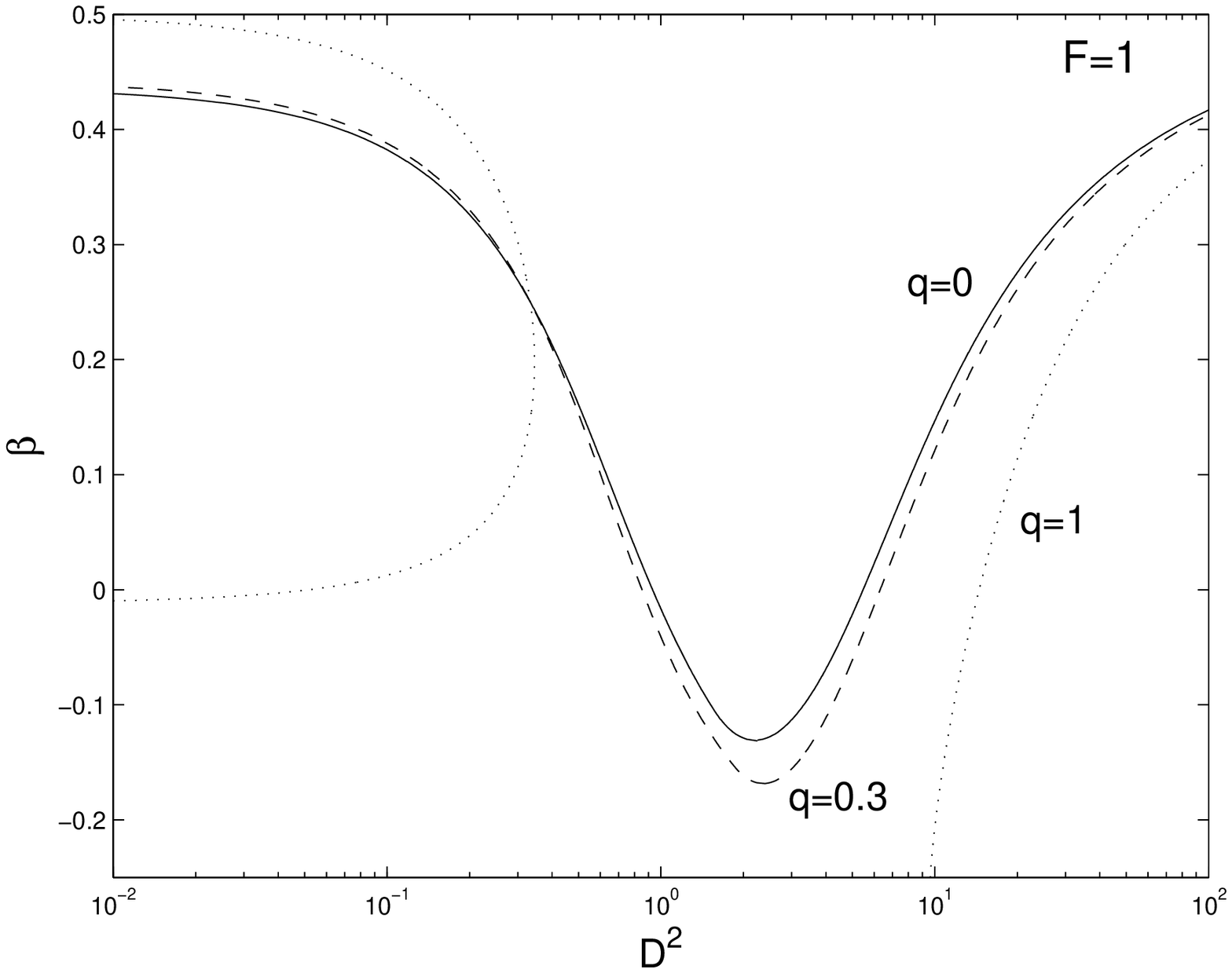}}
\subfigure[Stability boundary curves of $F$ versus
$D^2$ for $\beta=1/4$ discs with different values
of $q$ parameter.]{
\includegraphics[scale=0.42]{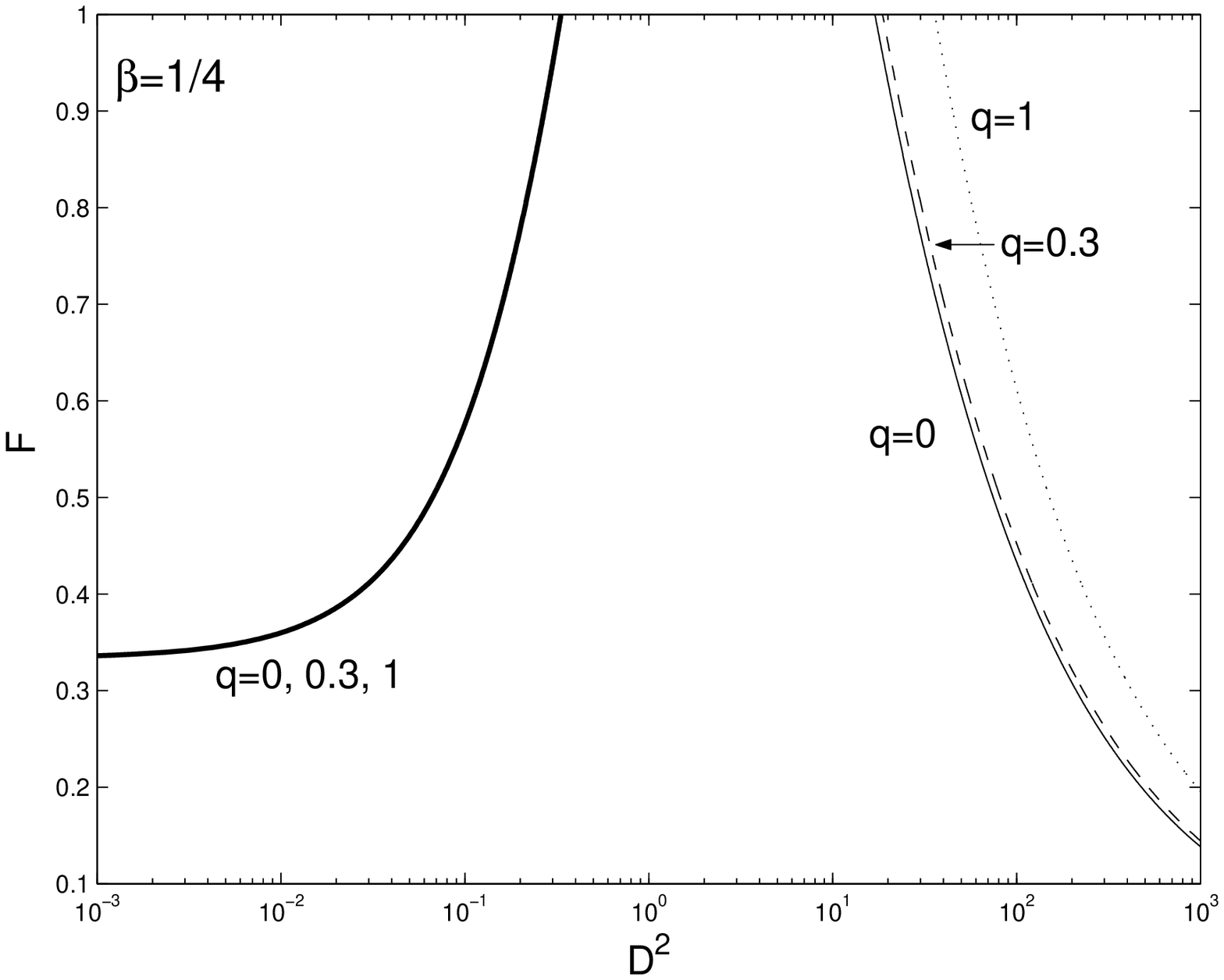}}
\caption{Variations of the stable $D^2$ range for (a) full discs
with $F=1$ and (b) $\beta=1/4$ discs. The solid, dashed and
dotted curves stand for $q=0$, $0.3$ and $1$, respectively. For
specific ranges of either $\beta\in(-1/4,1/2)$ [see panel (a)]
or $F\in(0,1]$ [see panel (b)], a horizontal line intercepts the
stability boundary curves at two values of $D^2$ and the stable
range of $D^2$ lies between the two intersection points. The $D^2$
range to the left of the smaller intersection point corresponds to
the collapse regime and the $D^2$ range to the right of the larger
intersection point corresponds to the ring-fragmentation regime.
Panel (a): For full discs with $F=1$, an enhancement of magnetic
field tends to suppress ring-fragmentation instabilities in general.
Meanwhile, an enhancement of magnetic field tends to suppress collapse
instabilities for $-1/4<\beta<1/4$ but to aggravate collapse
instabilities for $1/4<\beta<1/2$ and bears no effects on collapse
instabilities when $\beta=1/4$. For stability curves of moderate
magnetic field strengths (e.g., the solid and dashed curves), the
disc system becomes less stable for both instabilities when
$\beta$ decreases. As $\beta$ becomes sufficiently small, there
will be no stable range for $D^2$ [e.g., for $q=0$ without magnetic
field, this lower limit of $\beta$ is $\sim -0.130$ as derived in
Appendix D of Shen \& Lou (2004b)]. For a stronger magnetic field
(e.g., the dotted curve), the trend for ring-fragmentation
instabilities remains the same while the trend for collapse
instabilities is different depending on whether $\beta<1/4$ or not;
the collapse regime may disappear completely when $\beta$ becomes
small enough. Panel (b): For $\beta=1/4$ discs, an enhancement of
magnetic field tends to suppress the ring-fragmentation regime but
does not influence the collapse regime because the maximum of the
collapse branch remains always at $\xi=0$ such that the heavy solid
curve on the left actually represents a superposition of the three
$q$ curves. A decrease of $F$ (i.e., a more massive dark-matter
halo) tends to suppress both instabilities.  }
\end{figure}

\begin{figure}
\centering
\includegraphics[scale=0.42]{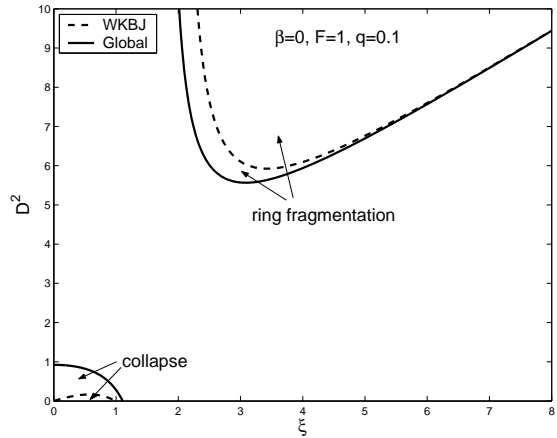}
\caption{A comparison of the global axisymmetric stability treatment
(solid curves) with the local WKBJ approach (dashed curves) for the
case of $\beta=0$, $F=1$ and $q=0.1$. The marginal stability curves
for the long-wavelength Jeans collapse and the short-wavelength ring
fragmentation regimes are denoted. For $\xi\gg 1$ (i.e., the
short-wavelength limit), the WKBJ approximation is well applicable,
while for small $\xi$ the results in the WKBJ approximation deviates
from those of the exact global treatment considerably (see Shen \&
Lou 2003, 2004b).}
\end{figure}


\subsubsection{Unaligned Logarithmic Spiral Cases with $m\ge 1$}

In parallel with the aligned $m\ge 1$ cases, we now construct
unaligned logarithmic spiral solutions from coplanar MHD
perturbation equations (\ref{stationary1})$-$(\ref{stationary4}).
For the density-potential pair $S\propto r^{-3/2+\hbox{i}\xi}$
and $V=-2\pi G{\cal N}_mrS$, we then have radial flow speed
$U\propto r^{-1/2+\beta+\hbox{i}\xi}$ and $z-$component specific
angular momentum $J\propto r^{1/2+\beta+\hbox{i}\xi}$. Coplanar
MHD perturbation equations $(\ref{stationary1})-(\ref{stationary4})$
can therefore be written in the forms of
\begin{equation}\label{spiralmge1}
\begin{split}
&m\Omega S+\frac{-1/2-\beta+\hbox{i}\xi}{r}\Sigma_0\hbox{i}U
+\frac{m\Sigma_0}{r^2}J=0\ ,\\
&m\Omega
\hbox{i}U+\frac{2\Omega}{r}J=\bigg(\frac{a^2}{r\Sigma_0}-2\pi
G{\cal N}_m\bigg) \bigg(-\frac{1}{2}+\hbox{i}\xi\bigg)S
-\frac{(1-4\beta)C_A^2}{2\Sigma_0r}S-\frac{C_A^2}{m\Omega r^2}
\bigg[\bigg(\frac{1}{2}-2\beta\bigg)\hbox{i}\xi
-\xi^2-m^2\bigg]\hbox{i}U\ ,\\
&m\Omega J+r(1-\beta)\Omega \hbox{i}U=m\bigg(2\pi G{\cal N}_m
-\frac{a^2}{r\Sigma_0}\bigg)rS+\frac{C_A^2(1-4\beta)}
{2\Omega r}\hbox{i}U\ ,\\
&Z=-\hbox{i}\xi\frac{B_0iU}{m\Omega r}\ ,
\qquad\qquad\qquad\qquad
R=\frac{B_0U}{\Omega r}\ .
\end{split}
\end{equation}
The first three equations of set
(\ref{spiralmge1}) immediately lead to
\begin{equation}\label{SUJspiral}
\begin{split}
a_1S+b_1\hbox{i}U+c_1J=0\ ,
\qquad\qquad
a_2S+b_2\hbox{i}U+c_2J=0\ ,
\qquad\qquad
a_3S+b_3\hbox{i}U+c_3J=0\ ,
\end{split}
\end{equation}
where coefficients $a_i$, $b_i$ and $c_i$
($i=1,2,3$) are explicitly defined by
\begin{equation}\label{abcspiral}
\begin{split}
&a_1\equiv m\Omega\ ,
\qquad\qquad\qquad
b_1\equiv\frac{-1/2-\beta+\hbox{i}\xi}{r}\Sigma_0\ ,
\qquad\qquad\qquad
c_1\equiv \frac{m\Sigma_0}{r^2}\ ,\\
&a_2\equiv \frac{(1-4\beta)C_A^2}{2\Sigma_0r}
+\bigg(-\frac{1}{2}+\hbox{i}\xi\bigg)
\bigg(2\pi G{\cal N}_m-\frac{a^2}{\Sigma_0 r}\bigg)\ ,
\qquad
b_2\equiv m\Omega+\frac{[(-2\beta+1/2)\hbox{i}\xi
-\xi^2-m^2]C_A^2}{m\Omega r^2}\ ,
\qquad
c_2\equiv \frac{2\Omega}{r}\ ,\\
&a_3\equiv -mr\bigg(2\pi G{\cal N}_m-\frac{a^2}{\Sigma_0 r}\bigg)\ ,
\qquad\qquad
b_3\equiv r(1-\beta)\Omega-\frac{(1-4\beta)C_A^2}{2\Omega r}\ ,
\qquad\qquad
c_3\equiv m\Omega\ .
\end{split}
\end{equation}
For nontrivial solutions of the set of homogeneous algebraic
equations (\ref{SUJspiral}), the determinant must vanish
\begin{equation}\label{spiral0}
a_1-\frac{b_1(a_2c_3-a_3c_2)}{b_2c_3-b_3c_2}
+\frac{c_1(a_2b_3-a_3b_2)}{b_2c_3-b_3c_2}=0\ .
\end{equation}

By using the set of expressions (\ref{abcspiral}) and applying
equilibrium conditions (\ref{PropEquilib}), we obtain the final
dispersion relation for stationary logarithmic spiral $m\ge 1$
MHD configurations in the form of a quadratic equation in terms
of $y\equiv D^2$, namely
\begin{equation}\label{spiral}
C_2y^2+C_1y+C_0=0\ ,
\end{equation}
where the coefficients $C_2$, $C_1$
and $C_0$ are explicitly defined by
\begin{equation}\label{spiralC2C1C0}
\begin{split}
&C_2\equiv {\cal H}_m\ ,\\
&C_1\equiv -\bigg[\frac{(3-4\beta)m^2-8\beta^2-
(4\xi^2-3)\beta+3\xi^2-1/4}{2(1+2\beta)}{F}
{\cal C}{\cal N}_m+m^2+\xi^2-\frac{7}{4}+7\beta\bigg]q^2
-{\cal A}_m(1-{F}{\cal C}{\cal N}_m)\ ,\\
&C_0\equiv -\frac{(1-4\beta )}{4(1+2\beta)}
\bigg[\frac{(-2m^2-2\xi^2-2\beta+1/2)}{(1+2\beta)}
{F}{\cal C}{\cal N}_m+1-4\beta\bigg]q^4
-\frac{(-2m^2-2\xi^2-2\beta+1/2)}{2(1+2\beta)}
(1-{F}{\cal C}{\cal N}_m)q^2\ ,
\end{split}
\end{equation}
with auxiliary parameters defined by
\begin{equation}\label{ABCDspiral}
\begin{split}
&{\cal A}_m\equiv m^2+\xi^2+\frac{1}{4}+2\beta\ ,\qquad
{\cal B}_m\equiv (1+2\beta)(m^2-2+2\beta)\ ,\qquad
{\cal C}\equiv \frac{(1+2\beta )}{2\beta{\cal P}_0}\ ,\qquad
{\cal H}_m\equiv F{\cal C}{\cal N}_m{\cal A}_m+{\cal B}_m\ \\
\end{split}
\end{equation}
that are the same as those in Shen \& Lou (2004b) for unaligned
logarithmic spiral cases. For a positive determinant
$\Delta\equiv C_1^2-4C_2C_0 >0$, the two real solutions
to quadratic equation (\ref{spiral}) are given by
\begin{equation}\label{Solspiral}
y_1=\frac{-C_1+\Delta^{1/2}}{2C_2}\
\ \qquad\qquad\hbox{ and }\qquad\qquad\
y_2=\frac{-C_1-\Delta^{1/2}}{2C_2}\ .
\end{equation}

Similar to $m\ge 1$ cases for aligned perturbations in the
limit of vanishing magnetic field (i.e., $q\rightarrow 0$),
equation (\ref{spiral}) reduces to
\begin{equation}\label{spiralNoM}
{\cal H}_my^2-{\cal A}_m(1-{F}{\cal C}{\cal N}_m)y=0\
\end{equation}
that leads to a nontrivial solution
\begin{equation}\label{spiralNoMSol}
D^2=y_1=\frac{(1-{F}{\cal C}{\cal N}_m){\cal A}_m}{{\cal H}_m}\ .
\end{equation}
Solution (\ref{spiralNoMSol}) is exactly the same result for the
single-disc case discussed in Section 3.2 of Shen \& Lou (2004b).
Another limiting regime of $\beta\rightarrow 0$ has been analyzed
by Lou (2002) for global stationary perturbation structures in a
single MSID.

\subsubsection{Force-Free Magnetized Discs with $\beta=1/4$}

For stationary spiral MHD configurations, we here analyze again
the $\beta=1/4$ case as a specific example of illustration. The
background equilibrium ring magnetic field is force-free and
inequality (\ref{restrt}) is satisfied. The determinant $\Delta$ of
quadratic equation (\ref{spiral}) remains always positive as shown
in Appendix A and hence two real solutions of $y=D^2$ to equation
(\ref{spiral}) exist. In parallel with the aligned $\beta=1/4$
case, we solve quadratic equation (\ref{spiral}) for given values
of parameter ${F}$ and display in diagrams the two branches of
solutions $y_1$ and $y_2$ ($y\equiv D^2$) versus $\xi$ variation.

As ${\cal B}_1<0$ for $m=1$ and when ${F}$ becomes sufficiently
small, there is a diverging point for $y_1\equiv D^2$ solution
where $C_2={\cal H}_1=0$. Since ${\cal H}_1$ increases monotonically
with increasing $\xi$ for $\xi\ge 0$ and attains its minimum value at
$\xi=0$ (${\cal H}_1$ is an even function of $\xi$; see Appendix C of
Shen \& Lou 2004b), the existence of such a critical point $\xi_c$
for a diverging $y\equiv D^2$ requires
\begin{equation}\label{criticalF}
{F}{\cal C}{\cal N}_1(0)(5/4+2\beta)+4\beta^2-1<0\ .
\end{equation}
For the specific $\beta=1/4$ case, this inequality (\ref{criticalF})
implies a critical ${F}$ value of ${F}_c=3/[7{\cal C}{\cal N}_1(0)]$
below which there exists one divergent point of $\xi_c$ where $y_1$
diverges. Numerically, we have determined ${F}_c\simeq0.6842$ [note
that this ${F}_c$ is of the same value as that of the aligned $m=1$
case because ${\cal N}_1(0)={\cal P}_1(1/4)$]. For selected $F$
values in the range of $0<{F}<{F}_c$, the corresponding critical
$\xi_c$ at which ${\cal H}_1=0$ can be numerically computed; for
example, we have $\xi_c=1.428$ for $F=0.5$.

It becomes much simpler for cases of $m\ge 2$ because $C_2$ remains
always positive and there is no diverging point for $D^2$ solutions.
Therefore, $y_1$ and $y_2$ remain always as the upper and lower
branches, respectively. For $m=1$ and $m=2$ cases with different
parameters of ${F}$ and $q$, we show in four panels $(a)-(d)$ of
Fig. 4 relevant curves of $y\equiv D^2$ versus $\xi$ variation to
illustrate the basic features.
To be more explicit for the case of $\beta=1/4$
in quadratic equation (\ref{spiral}), we write
\begin{equation}\!\!\!\label{beta0d25s}
\begin{split}
&\qquad\qquad\bigg
[{F}{\cal C}{\cal N}_m\bigg(m^2+\xi^2+\frac{3}{4}\bigg)
+\frac{3}{2}\bigg(m^2-\frac{3}{2}\bigg)\bigg]D^4
-\bigg[(m^2+\xi^2)\bigg(\frac{2}{3}
{F}{\cal C}{\cal N}_m+1\bigg)q^2\\
&\hbox{ }\qquad\qquad\qquad\qquad\qquad\qquad\qquad
+\bigg(m^2+\xi^2+\frac{3}{4}\bigg)
(1-{F}{\cal C}{\cal N}_m)\bigg]D^2
+\frac{2}{3}(m^2+\xi^2)(1-{F}{\cal C}{\cal N}_m)q^2=0\ ,
\end{split}
\end{equation}
where auxiliary parameters ${\cal N}_m$, ${\cal A}_m$,
${\cal B}_m$, ${\cal C}$ and ${\cal H}_m$ are all
evaluated for the special value of $\beta=1/4$.

\begin{figure}
\centering \subfigure[$m=1$, $\beta=1/4$, ${F}=1$, $q=0,\
1$]{
\includegraphics[scale=0.42]{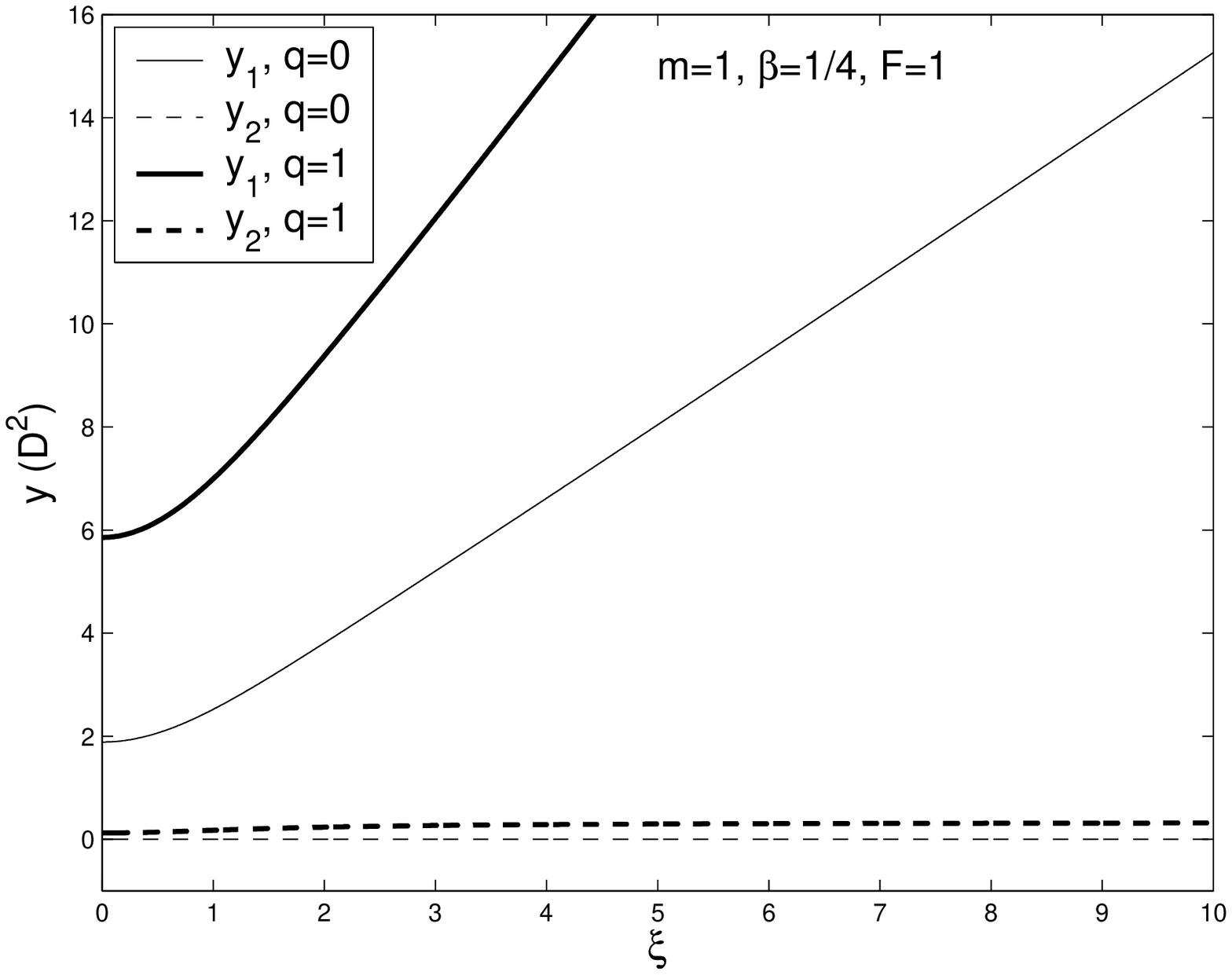}}
\subfigure[$m=1$, $\beta=1/4$, ${F}=0.5$, $q=0,\ 1$]{
\includegraphics[scale=0.42]{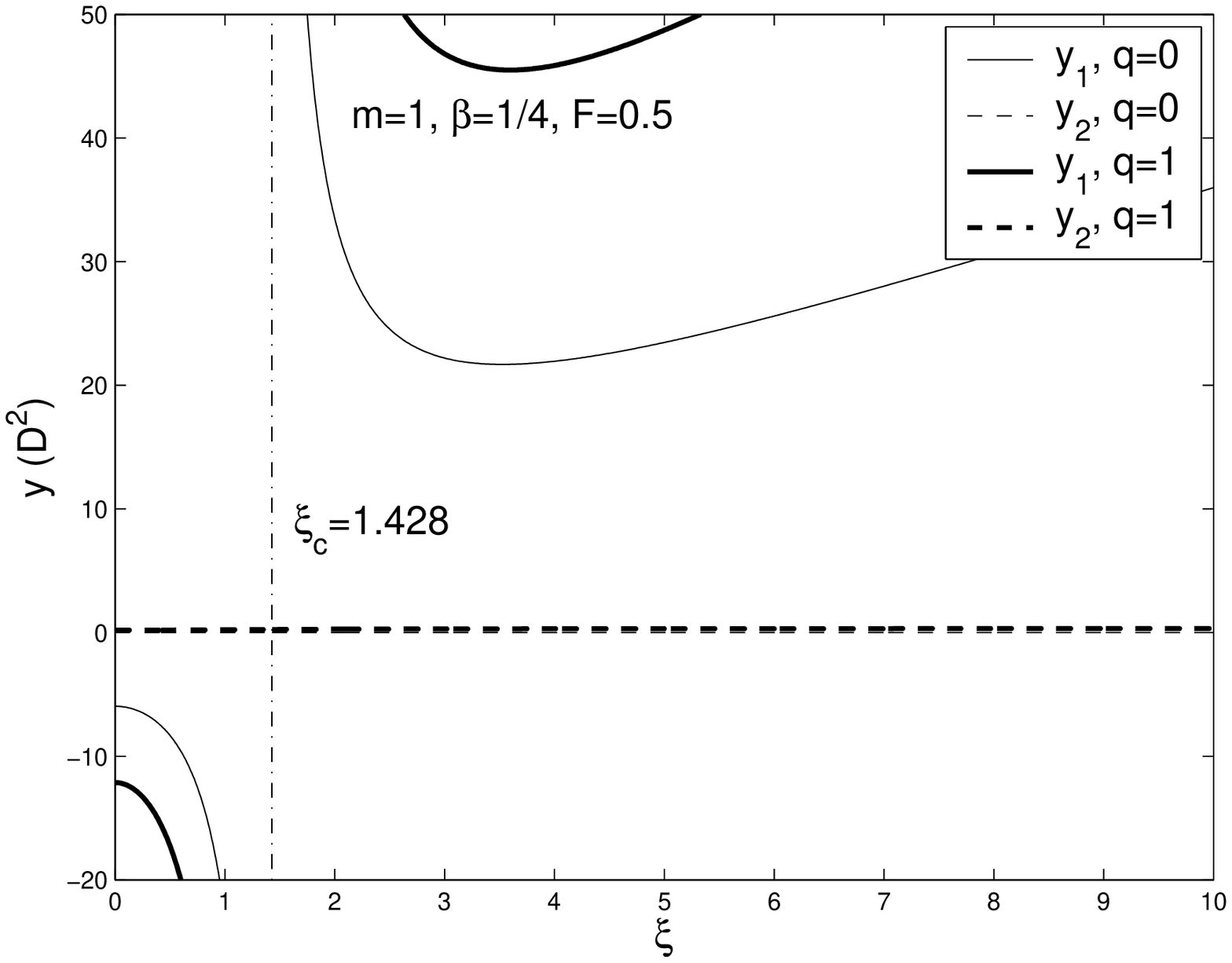}}
\subfigure[$m=2$, $\beta=1/4$, ${F}=1$, $q=0,\ 1$]{
\includegraphics[scale=0.42]{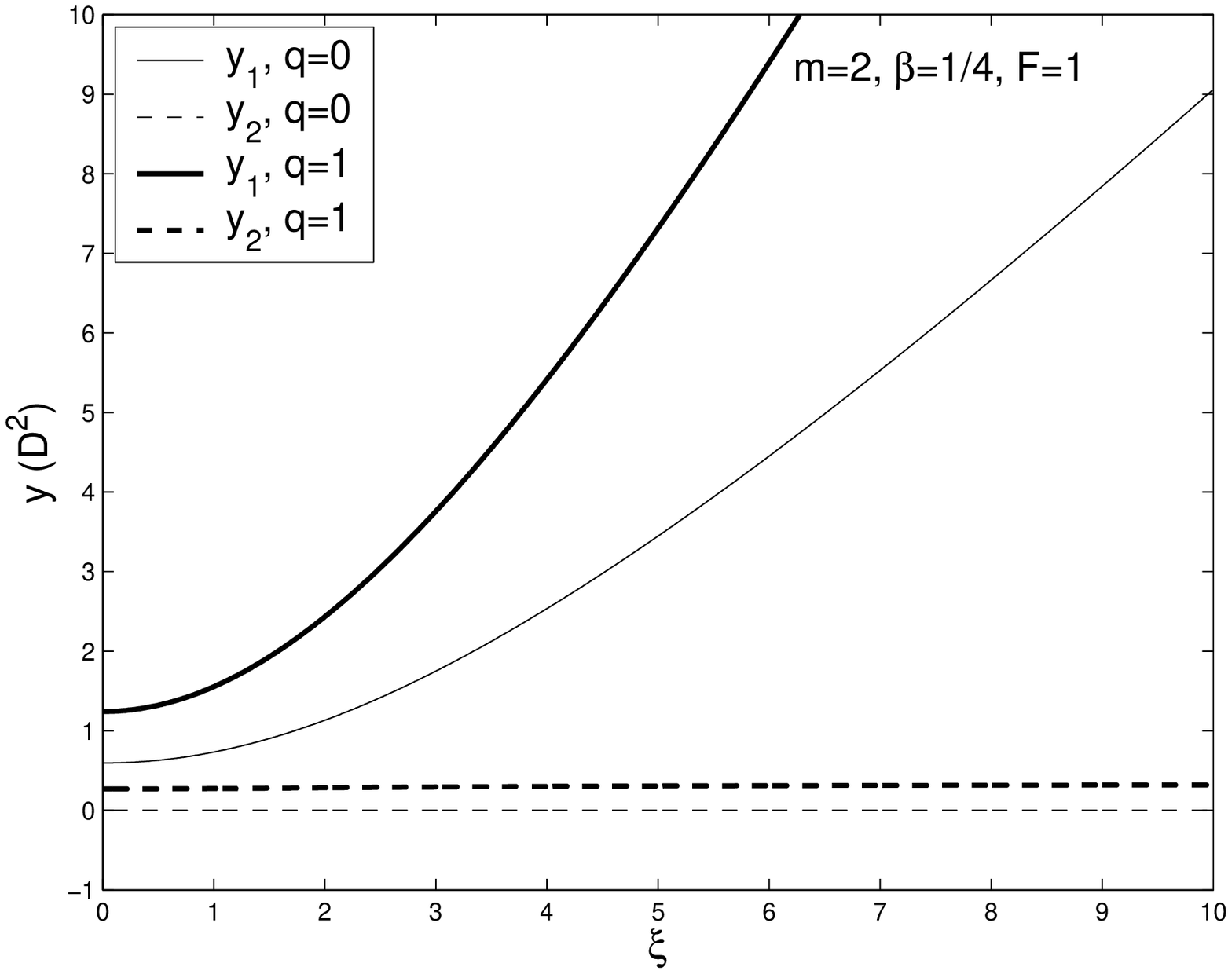}}
\subfigure[$m=2$, $\beta=1/4$, ${F}=0.5$, $q=0,\ 1$]{
\includegraphics[scale=0.42]{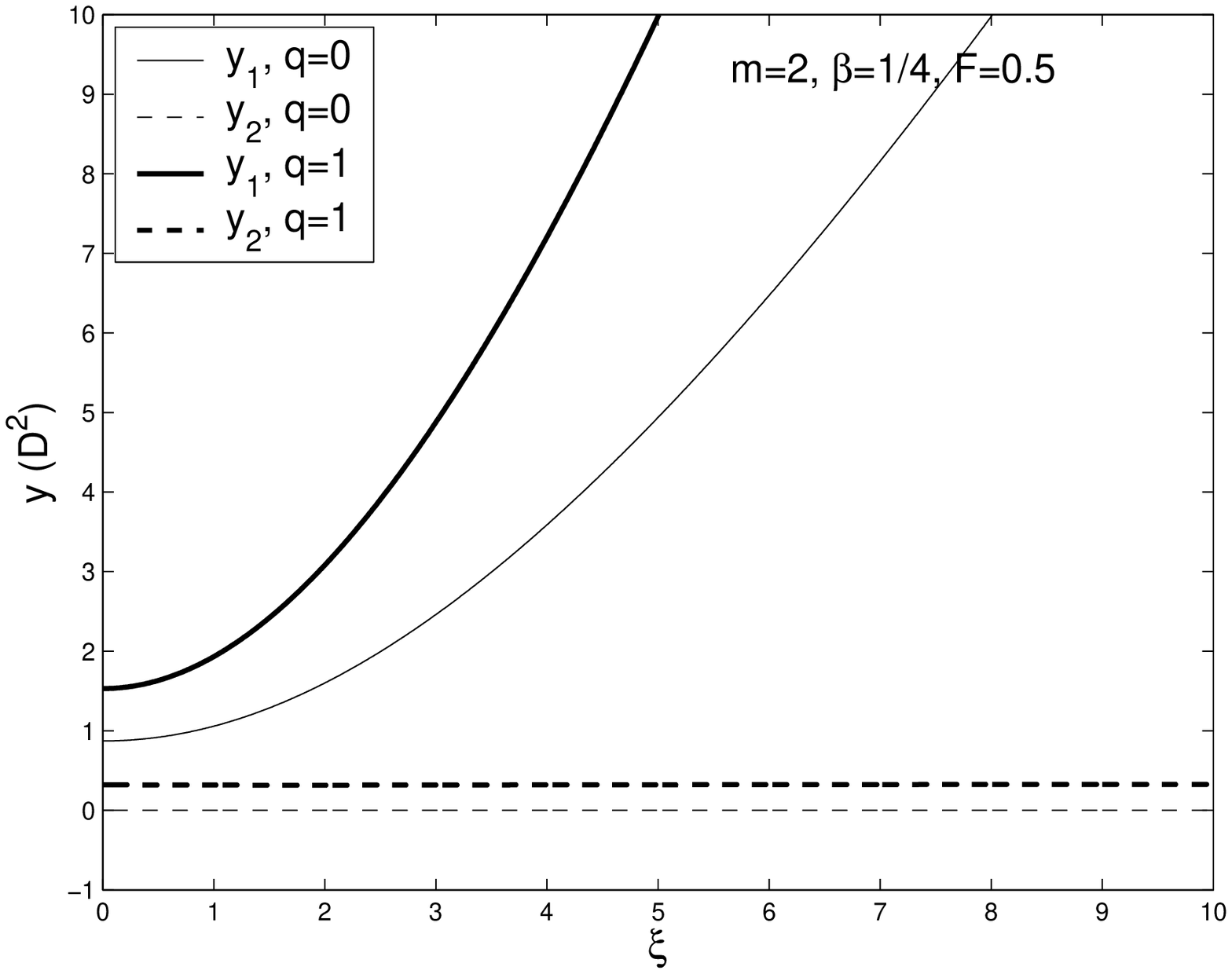}}
\caption{Two $D^2$ solution branches $y_1$ and $y_2$ of
quadratic equation (\ref{spiral}) for $m=1,\ 2$, $\beta=1/4$,
$F=1,\ 0.5$ and $q=0,\ 1$. The $y_1$ and $y_2$ branches are
plotted in solid and dashed lines, respectively. Parameters
adopted are shown in panels $(a)-(c)$.
}
\end{figure}

\section{Perturbation Phase Relationships and Angular Momentum Transfer}

For multi-wavelength observations of nearby spiral galaxies (Sofue
et al. 1986; Kronberg 1994; Beck et al. 1996), it is possible to
identify large-scale spatial phase relationships of spiral patterns
using various observational diagnostics (e.g., Mathewson et al. 1972;
Visser 1980a, b; Neininger 1992; Beck \& Hoernes 1996; Frick et al.
2000, 2001; Ferri\`ere 2001; Lou et al. 2002). For this reason, we
derive below spatial phase relationships for various coplanar MHD
perturbation variables, keeping in mind of potential applications to
magnetized spiral galaxies (e.g., Fan \& Lou 1996; Lou \& Fan 1998a).

\subsection{General Perturbation Phase Relationships
for both Aligned and Spiral Cases}

Here we systematically examine the spatial phase relationships among
enhancements of gas surface mass density, magnetic field and velocity
perturbations by analysing the relationships among $S$, $Z$, $R$, $J$
and $iU$ for aligned and logarithmic spiral cases, separately. Starting
from the set of homogeneous algebraic equations (\ref{SUJ}) for the
aligned cases, we obtain the relationships among $S$, $U$ and $J$ in
the following, namely
\begin{equation}\label{RelSUJ}
\frac{\hbox{i}U}{S}=\frac{c_3a_1-c_1a_3}{c_1b_3-c_3b_1}\ , \
\qquad\qquad\qquad
\frac{J}{S}=\frac{a_2b_1-a_1b_2}{c_1b_2-c_2b_1}\ ,\
\end{equation}
where $a_i$, $b_i$ and $c_i$ ($i=1,2,3$) are defined by
expressions (\ref{abc}). By using definitions (\ref{abc})
and equations (\ref{alignedmge1}), we obtain
\begin{equation}\label{RelSURJZ}
\begin{split}
&\frac{\hbox{i}U}{S}=\frac{m\Omega r}{\Sigma_0}\Theta_{A}\ ,
\quad\qquad
\frac{Z}{S}=-\frac{(1/2-2\beta)B_0}{\Sigma_0}\Theta_{A}\ ,
\quad\qquad
\frac{\hbox{i}R}{S}=\frac{mB_0}{\Sigma_0}\Theta_{A}\ ,
\quad\qquad
\frac{J}{S}=-\frac{\Omega r^2}{\Sigma_0}(1-3\beta\Theta_{A})\
\end{split}
\end{equation}
for the aligned cases, where the common dimensionless
real factor $\Theta_{A}$ is defined below by
\begin{equation}\label{ThetaA}
\Theta_{A}\equiv\frac{a^2-2\pi G{\cal P}_m\Sigma_0r-\Omega^2r^2}
{(1/2-2\beta)C_A^2-(1+2\beta)\Omega^2r^2}\ .
\end{equation}

Similarly, by using the set of homogeneous algebraic equations
(\ref{SUJspiral}), definitions (\ref{abcspiral}) and equations
(\ref{spiralmge1}), we readily derive for the cases of global
stationary logarithmic spiral configurations
\begin{equation}\label{RelSURJZspiral}
\begin{split}
&\frac{\hbox{i}U}{S}=\frac{m\Omega r}{\Sigma_0}\Theta_{S}\ ,
\qquad\quad \frac{Z}{S}=-\frac{\hbox{i}\xi
B_0}{\Sigma_0}\Theta_{S}\ , \qquad\quad
\frac{\hbox{i}R}{S}=\frac{mB_0}{\Sigma_0}\Theta_{S}\ , \qquad\quad
\frac{J}{S}=-\frac{\Omega r^2}{\Sigma_0}
\bigg[1+\bigg(-\frac{1}{2}-\beta+\hbox{i}\xi\bigg)\Theta_{S}\bigg]\ ,
\end{split}
\end{equation}
where the common dimensionless complex
factor $\Theta_{S}$ is defined below by
\begin{equation}\label{ThetaS}
\Theta_{S}\equiv\frac{a^2-2\pi G{\cal N}_m\Sigma_0r-\Omega^2r^2}
{(1/2-2\beta)C_A^2-(3-2\hbox{i}\xi )\Omega^2r^2}\ .
\end{equation}

The two sets of algebraic equations (\ref{RelSURJZ}) and
(\ref{RelSURJZspiral}) are the most general phase relationships for
aligned and logarithmic spiral cases, respectively. Once a $\beta$
value is specified and a particular $D^2$ for a stationary solution
inserted, we immediately obtain the phase relationship between any
two perturbation variables. As an example of illustration, we
examine again the special $\beta=1/4$ case and determine the relevant
phase relationships in the next subsection.

\subsection{The $\beta=1/4$ Force-Free Case}

When $\beta=1/4$, the results of phase relations derived in the
last subsection 4.1 can be simplified considerably as all terms
involving the factor $1/2-2\beta$ vanish. Meanwhile, inequality
(\ref{restrt}) is automatically satisfied. The only requirement
for a physical $D^2$ solution is simply $D^2>0$. As already noted
in subsections 3.1.3 and 3.2.3, there are two branches of $D^2$
solutions. We insert relevant physical $D^2$ solutions in
equations (\ref{RelSURJZ}) and (\ref{RelSURJZspiral}) to determine
phase relationships among perturbation variables. In particular,
we are interested in the phase relationship between the perturbed
surface mass density and the perturbed azimuthal magnetic field.
We now analyze below this mass-magnetic field phase relationships
for aligned and spiral cases, separately. Some illustrative
examples are shown in Fig. 5 and Fig. 6 to provide more direct
visual impressions.

\subsubsection{Aligned Cases}

For aligned $\beta=1/4$ cases, we readily find $Z=0$ which
means the azimuthal magnetic field remains unchanged. From
equation (\ref{RelSURJZ}), we derive
\begin{equation}\label{SR}
\hbox{i}R/S\propto 3D^2/2-1+{F}{\cal C}{\cal P}_m(1+D^2)
=({F}{\cal C}{\cal P}_m +3/2)D^2-(1-{F}{\cal C}{\cal P}_m)\ .
\end{equation}
Meanwhile, equation (\ref{align0d25})
can be rewritten in the form of
\begin{equation}\label{q2aligned}
\frac{m^2q^2(1+2{F}{\cal C}{\cal P}_m/3)}
{(m^2+3/4)[F{\cal C}{\cal P}_m+3(m^2-3/2)/(2m^2+3/2)] }
=\frac{D^2\{D^2-(1-F{\cal C}{\cal P}_m)/
[F{\cal C}{\cal P}_m+3(m^2-3/2)/(2m^2+3/2)]\} }
{D^2-(1-F{\cal C}{\cal P}_m)/(F{\cal C}{\cal P}_m+3/2) }\ ,
\end{equation}
that contains two distinct $D^2$ solutions according to subsection
3.1.3, namely, the plus-sign solution $y_1$ and the minus-sign
solution $y_2$ in the solution form (\ref{Solalign}), respectively.
As shown in Appendix B, we emphasize that the plus-sign solution
$y_1$ (if positive and thus physical) makes
$D^2-(1-F{\cal C}{\cal P}_m)/(F{\cal C}{\cal P}_m+3/2)>0$ and
the minus-sign solution $y_2$ makes
$D^2-(1-F{\cal C}{\cal P}_m) /(F{\cal C}{\cal P}_m+3/2)<0$ on
the right-hand side of equation (\ref{q2aligned}), respectively.
Therefore, the two branches of $D^2$ solutions (if positive and
thus physical) will give either $\hbox{i}R/S>0$ for the plus-sign
$y_1$ solution or $\hbox{i}R/S<0$ for the minus-sign $y_2$ solution.

\subsubsection{Cases of Stationary Logarithmic Spiral Configurations}

For spiral cases with $\beta=1/4$, we readily obtain from
equations (\ref{RelSURJZspiral}) and (\ref{ThetaS}) that
\begin{equation}\label{thetaS}
\Theta_{S}\propto [({F}{\cal C}{\cal N}_m +3/2)D^2-(1-{\cal
F}{\cal C}{\cal N}_m)](3+2\hbox{i}\xi)\ .
\end{equation}
Following the same procedure, we can show that the two branches
of $D^2$ solutions, namely, the plus-sign solution $y_1$ and the
minus-sign solution $y_2$ given by equation (\ref{Solspiral}),
if positive and thus physical, correspond to the factor
$(F{\cal C}{\cal N}_m+3/2)D^2-(1-F{\cal C}{\cal N}_m)$ being
positive and negative, respectively (see Appendix B for a
detailed proof). We now examine the phase relationship between
coplanar perturbation enhancements of surface mass density and
magnetic field.

In the short-wavelength limit (i.e., $\xi\gg 1$ with a logarithmic
spiral pattern tightly wound), we have $Z/S\propto\pm 1$ and
$R/S\propto\pm 1$ where the plus- and minus-signs correspond to
the $y_1$ and $y_2$ solutions, respectively. Therefore, the phase
relations between perturbation enhancements of surface mass density
and magnetic field are in-phase for the $y_1$ solution (if positive
and physical) and out-of-phase for the $y_2$ solution. For $m\ge 2$,
$y_1$ remains always greater than $y_2$. In contexts of magnetized
spiral galaxies, these solution properties in the WKBJ regime are
consistent with the physical identifications of the $y_1$ mode
with a stationary fast MHD density wave and the $y_2$ mode with a
stationary slow MHD density wave, respectively (Fan \& Lou 1996;
Lou \& Fan 1998a; Lou \& Fan 2002). It is now possible to model
magnetized spiral galaxies in terms of coplanar stationary fast and
slow MHD density waves with more general rotation curves including
the case of a flat rotation curve for a magnetized singular
isothermal disc (Shu et al. 2000; Lou 2002; Lou \& Fan 2002; Lou
\& Zou 2004a, b; Lou \& Wu 2004).

In the long-wavelength limit (i.e., $\xi\ll 1$ with a logarithmic
spiral pattern widely open), we have $Z/S\propto\mp\hbox{i}$ and
$R/S\propto\mp\hbox{i}$ where the minus- and plus-signs correspond
to the $y_1$ and $y_2$ solutions, respectively. Therefore, the
coplanar perturbation enhancement of magnetic field is either
ahead of or lagging behind the perturbation enhancement of surface
mass density by a phase difference of $\sim\pi/2$.

For intermediate radial wavelengths, the phase difference between
perturbation enhancements of surface mass density and magnetic
field in a logarithmic spiral pattern can be readily determined
for given specific conditions.

On the basis of our model analysis here, it appears that except
for global stationary fast MHD density waves in the extreme WKBJ
regime, magnetic and gas spiral arms tend to be phase shifted
relative to each other for a global stationary logarithmic spiral
pattern for a much broader set of rotation curves including the
special case of a flat rotation curve.


\begin{figure}
\centering\subfigure[Fractional perturbed surface mass density
gray-scale plot as a spatial phase reference for either fast
or slow MHD density wave modes]{
\includegraphics[scale=0.42]{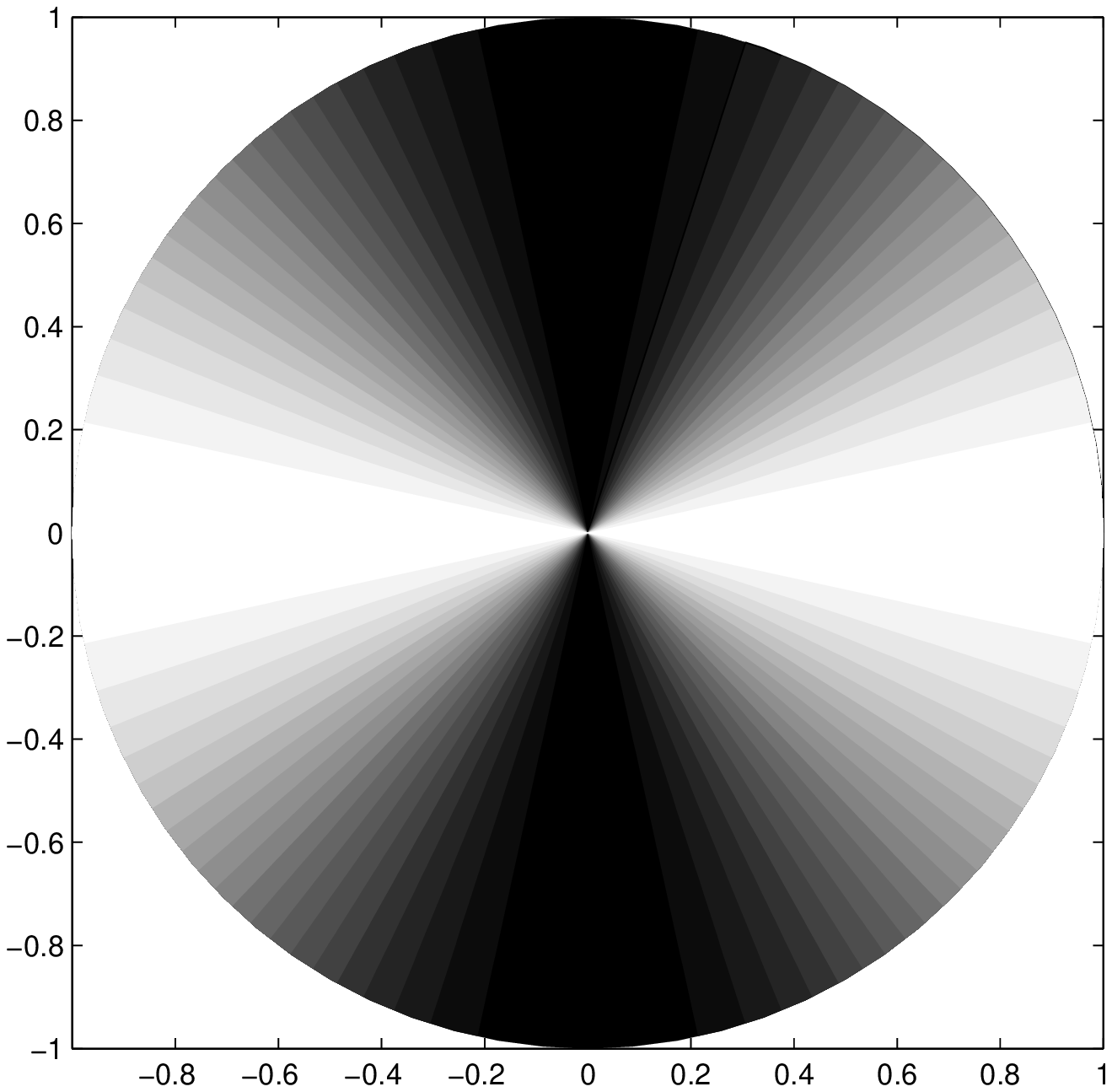}}
\subfigure[Fractional perturbed radial magnetic field
in a gray-scale plot for the fast MHD density wave mode]{
\includegraphics[scale=0.42]{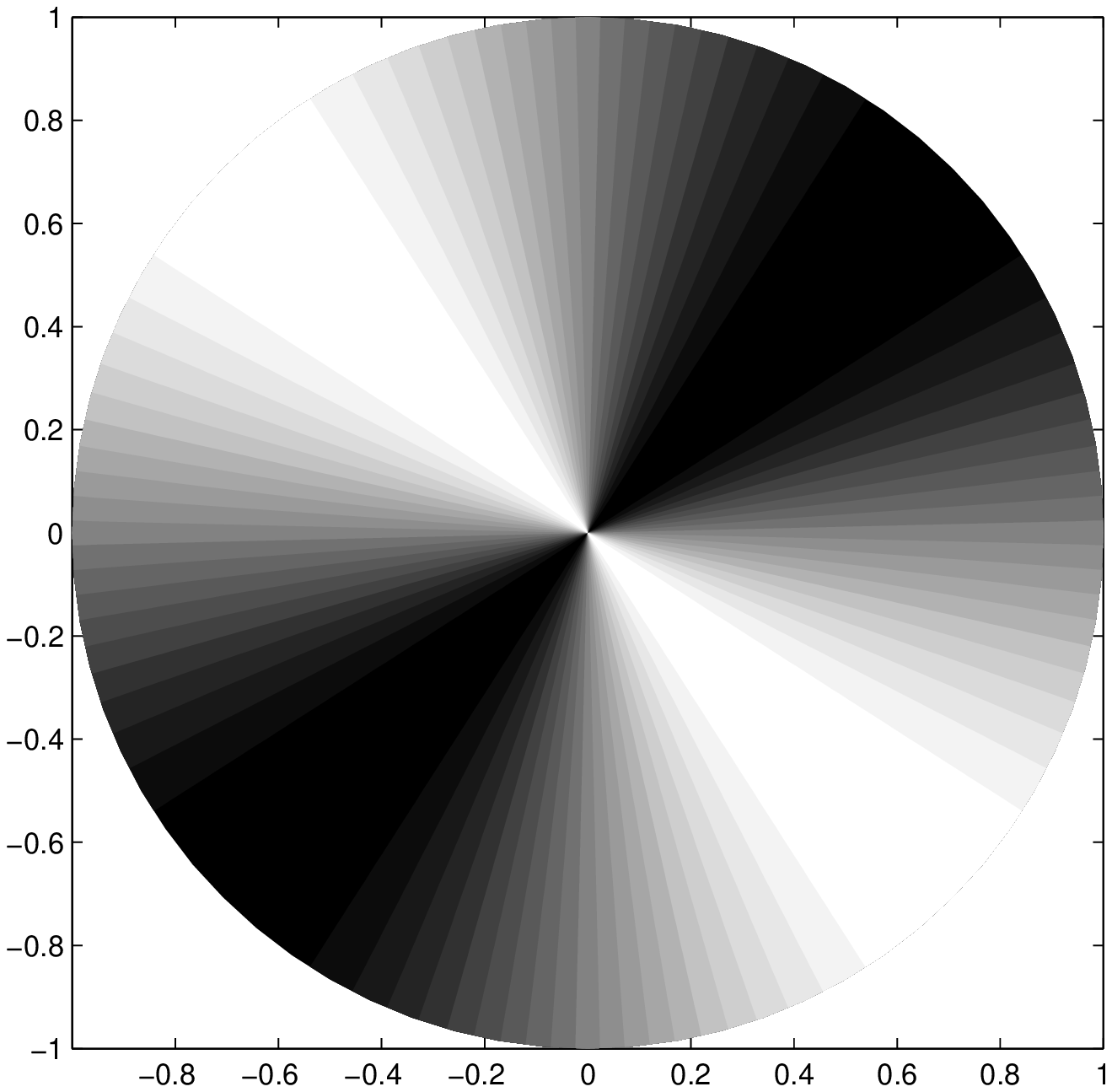}}
\subfigure[Fractional perturbed radial magnetic field in
a gray-scale plot for the slow MHD slow density wave mode]{
\includegraphics[scale=0.42]{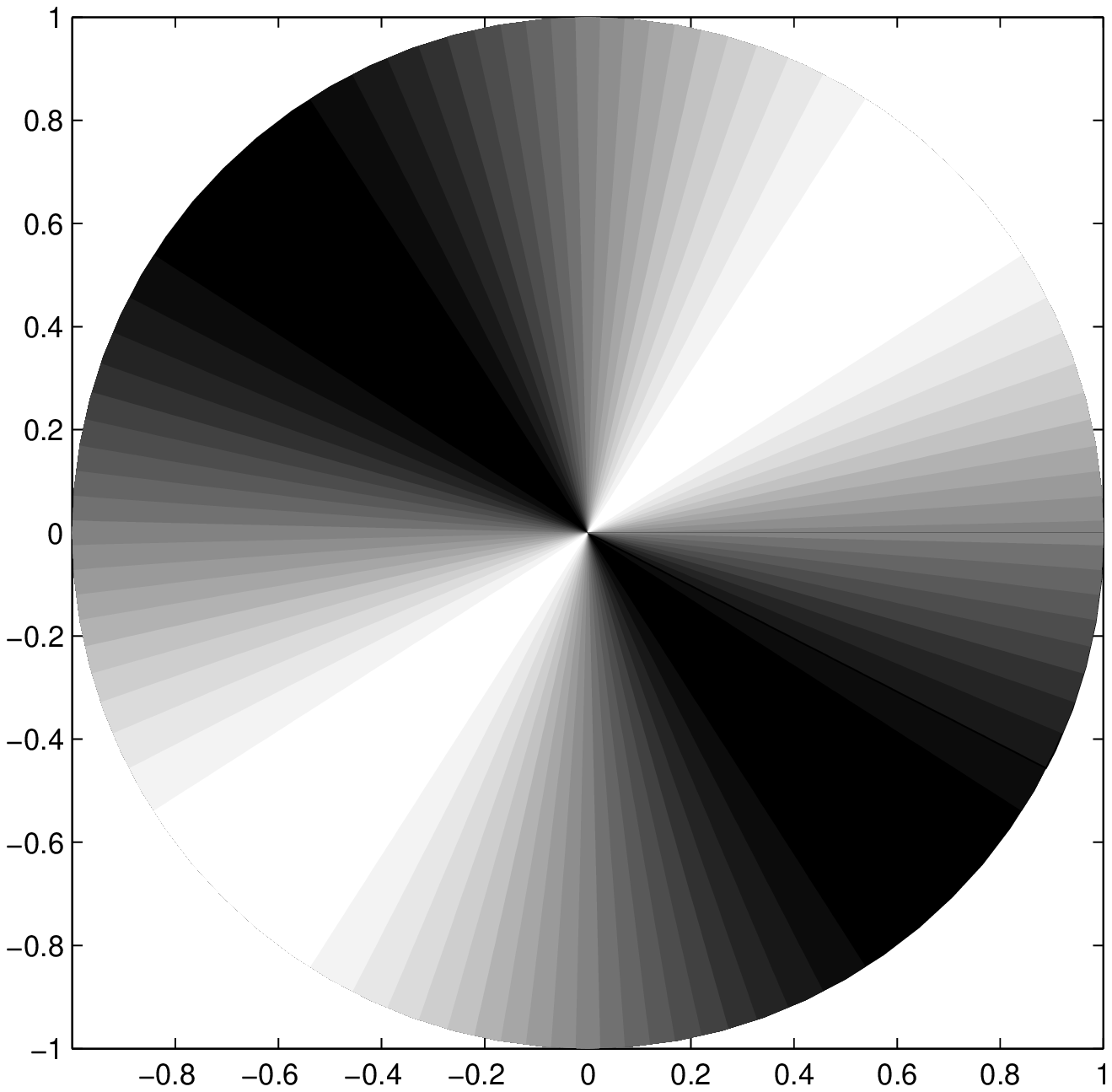}}
\caption{Spatial phase relationship between the perturbed
surface mass density and the radial magnetic field for the aligned
$m=2$ bar-like case for the specific case of $\beta=1/4$. Shown in
the figures above are the fractional perturbed surface mass density
and magnetic field contours (i.e., perturbation variables divided
by the radial scalings of their equilibrium quantities with
$r^{-3/2}$ for the surface mass density and $r^{-1}$ for the
azimuthal magnetic field). The brightest regions correspond to
the strongest density enhancement. Panel (a) is for the perturbed
relative surface mass density; panels (b) and (c) are for the
perturbed relative radial magnetic fields for the fast and slow
MHD perturbations respectively. According to equation (\ref{SR}),
there is a phase shift of $\mp\pi/2$ for the global stationary
fast and slow MHD density wave modes. Since $m=2$, the pattern
appears a rotation of $\mp\pi/4$. }
\end{figure}

\begin{figure}
\centering \subfigure[Fractional perturbed mass density in a
gray-scale plot as a spatial phase reference for either fast
or slow MHD density wave modes]{
\includegraphics[scale=0.42]{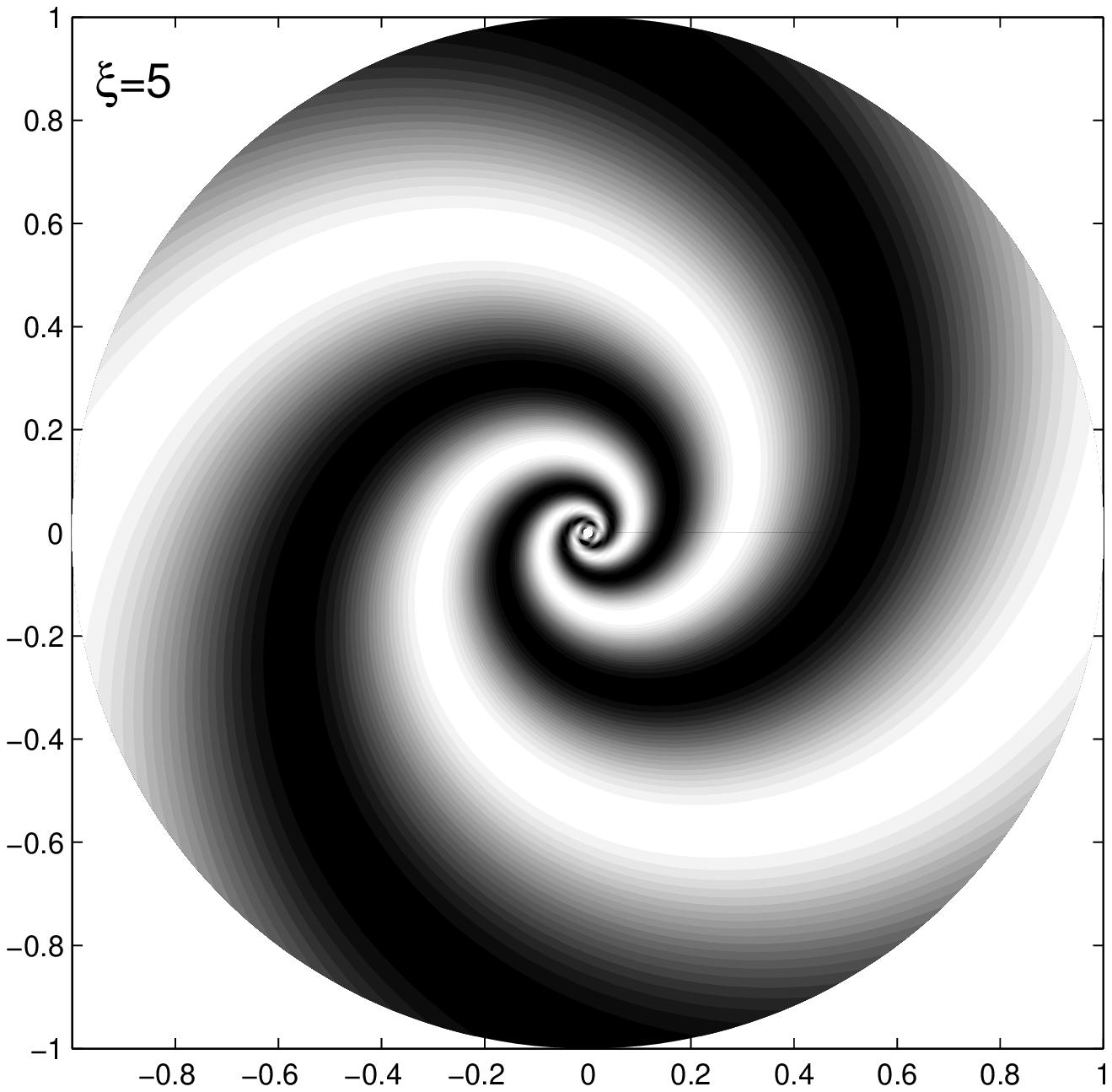}}
\subfigure[Fractional perturbed azimuthal magnetic field
in a gray-scale plot for the fast MHD density wave mode]{
\includegraphics[scale=0.42]{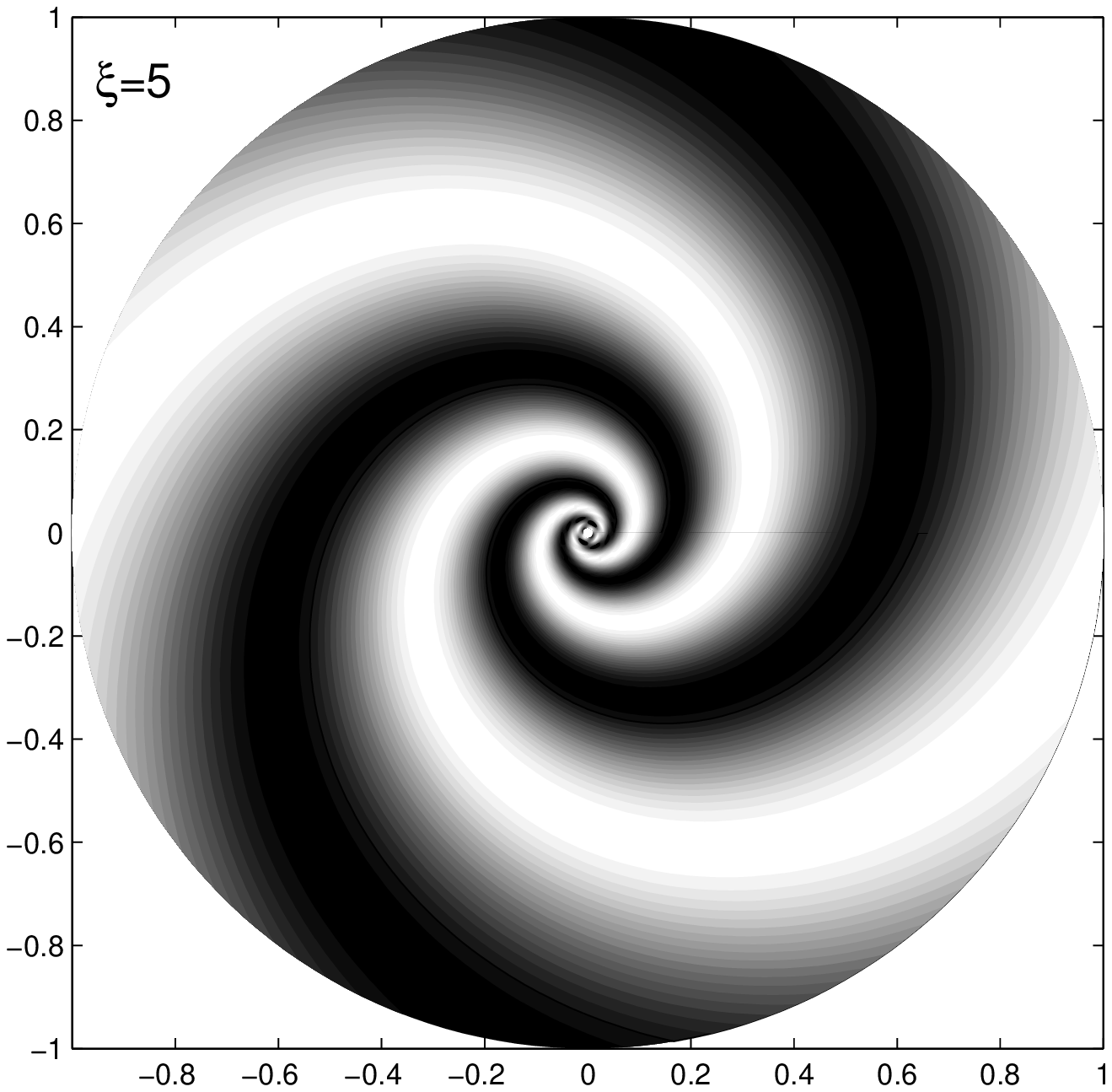}}
\subfigure[Fractional perturbed azimuthal magnetic field
in a gray-scale plot for the slow MHD density wave mode]{
\includegraphics[scale=0.42]{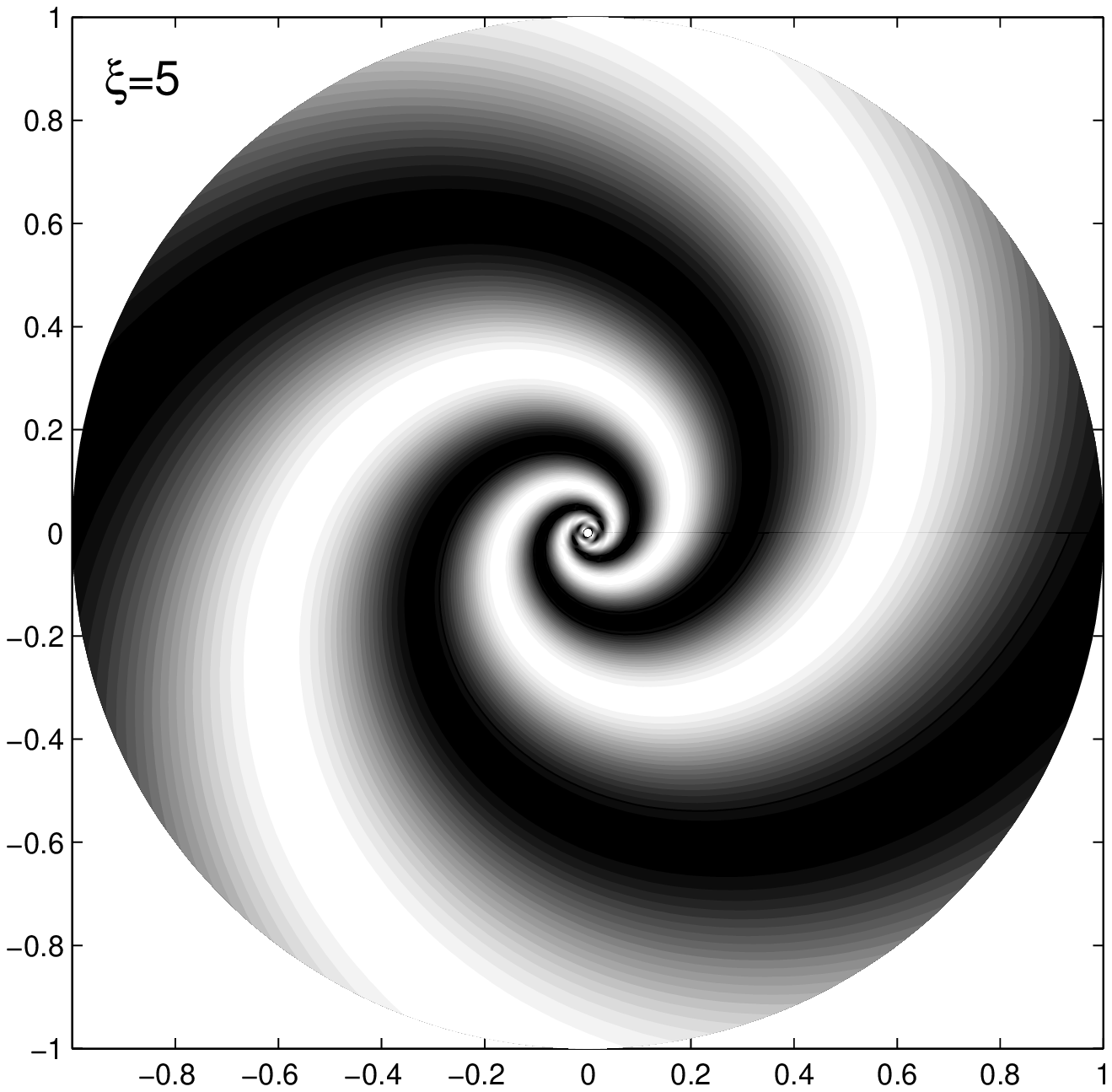}}
\caption{Phase relationship between the perturbed surface mass
density and the azimuthal magnetic field for the logarithmic spiral
$m=2$ case (an effective radial wavenumber $\xi=5$) for the specific
case of $\beta=1/4$. Shown in above three panels in order are the
fractional perturbed surface mass density and azimuthal magnetic
field gray-scale plots (i.e., divided by the radial scaling of their
respective equilibrium quantities with $r^{-3/2}$ for the surface mass
density and $r^{-1}$ for the azimuthal  magnetic field). The brightest
regions correspond to the most density enhancement. Panel (a) is for
the perturbed relative surface mass density; panels (b) and (c) are
for the perturbed azimuthal magnetic fields for the fast and slow MHD
perturbations, respectively. According to equations (\ref{thetaS}) and
(\ref{RelSURJZspiral}), there is a phase shift of $\sim -0.093\pi$
($\sim 0.907\pi$) for the fast (slow) MHD perturbation mode. Since
$m=2$, the pattern appears a rotation of half of the phase shift.
For the fast MHD density wave mode, the magnetic field enhancement
roughly follows the spirals of surface mass enhancement. For the
slow MHD density wave mode, the density spiral arms and the magnetic
spiral arms largely interlace with each other. For $\xi\gg 1$, the
phase pattern relationships become those in the usual WKBJ regime
(e.g., Fan \& Lou 1996; Lou \& Fan 1998a). }
\end{figure}

\subsection{Angular Momentum Conservation}

One can readily show that for aligned global stationary MHD
perturbation configurations, there is no net radial angular
momentum flux as aligned stationary MHD density waves do not
propagate radially (Shu et al. 2000; Lou 2002). We here briefly
discuss the process of angular momentum transfer in association
with the stationary logarithmic spiral MHD modes obtained in
this paper. According to the analyses of Lynden-Bell \& Kalnajs
(1972) and Fan \& Lou (1999), the total angular momentum flux
carried by coplanar MHD density waves should contain three
separate contributions, namely, the advective transport flux
$\Lambda^{A}$ defined by
\begin{equation}\label{adveangm}
\Lambda^{A}\equiv r^2\Sigma_0\int_0^{2\pi}
d\theta\Re(u_1)\Re(j_1/r)\ ,
\end{equation}
the gravity torque flux $\Lambda^{G}$ defined by
\begin{equation}\label{gravangm}
\Lambda^{G}\equiv\frac{1}{4\pi G}
\int_0^{2\pi}d\theta\int_{-\infty}^{\infty}dzr
\bigg[\frac{\partial\Re(\phi_1)}
{\partial\theta}\bigg]\bigg[\frac{\partial\Re(\phi_1)}
{\partial r}\bigg]\
\end{equation}
and the magnetic torque flux $\Lambda^{B}$ defined by
\begin{equation}\label{magnangm}
\Lambda^{B}\equiv -\frac{r^2}{4\pi}\int_0^{2\pi}
d\theta\int_{-\infty}^{\infty}dz\Re(b_r)\Re(b_\theta)\ .
\end{equation}

The perturbed three-dimensional gravitational potential
$\phi_1(r,\theta,z)$ associated with a Fourier component of a
coplanar logarithmic spiral perturbation in surface mass density
$\Sigma_1=\sigma r^{-3/2+\hbox{i}\xi}e^{-im\theta}$ is then
\begin{equation}\label{3dphi1}
\phi_1(r,\theta,z)=-2\pi G\sigma
e^{-\hbox{i}m\theta}\int_0^{\infty}dke^{-k|z|}J_m(kr)
\int_0^{\infty}r_a^{-1/2+\hbox{i}\xi}J_m(kr_a)dr_a\ ,
\end{equation}
where $J_m(u)$ is the cylindrical Bessel function of order
$m$ with an argument $u$ (e.g., Binney \& Tremaine 1987).
In fact, integral (\ref{3dphi1}) gives a simple form of
Kalnajs function in the disc plane at $z=0$ with the real
part of $\phi_1(r,\theta,z)$ explicitly given by
\begin{equation}\label{rmphi}
\Re(\phi_1)=-2\pi G\sigma\int_0^{\infty}dke^{-k|z|}J_m(kr)
\int_0^{\infty}r_a^{-1/2}\cos(m\theta-\xi\ln r_a)J_m(kr_a)dr_a\ .
\end{equation}
It is fairly straightforward to show that $\Lambda^{A}$ and
$\Lambda^{B}$ are independent of $r$ where the phase relationships
for stationary logarithmic spiral cases derived in Section 4.1
have been use. While mathematically tedious, one can further show
that $\Lambda^{G}$ is also independent of $r$ (see Appendix C).
Therefore, the total angular momentum flux $\Lambda^A+\Lambda^G+
\Lambda^B$ remains constant at all radii, a natural result of
angular momentum conservation given that perturbation surface mass
density falls off as $r^{-3/2}$. Therefore, in order to maintain a
stationary logarithmic spiral pattern, there must be a source or a
sink at the disc center.
The perturbation energy flux, however, is zero at all radii for
the stationary coplanar MHD configurations [see equation (4.17)
of Fan \& Lou (1999) in the WKBJ regime]. The mass flux is also
zero at all radii since the first-order mass flux
$\int_0^{2\pi}\Sigma_0u_1rd\theta=0$, and the second-order mass flux
$\int_0^{2\pi}(\Sigma_1u_1+\Sigma_0u_2)rd\theta=0$ [where $u_2$ is
the second-order perturbed radial velocity, this equation holds due
to the second-order mass conservation (Fan \& Lou 1999)]. Therefore
our model result differs from that of Spruit (1987) for stationary
logarithmic spiral shocks, where only the gravity of the central
object is considered. In Spruit's model, there is a net mass accretion
through the disc and an effective viscosity parameter ``$\alpha$''
(Shakura \& Sunyaev 1973) may be estimated.

Physically, the central source of angular momentum required to
sustain global stationary logarithmic spiral patterns may be
plausibly identified with a fast spinning supermassive black hole
(SMBH) in the nucleus of a spiral galaxy. The Kerr spacetime
associated with such a spinning SMBH excites and sustains MHD
density waves in the surrounding disc through a long-range
gravitational interaction. For trailing spiral patterns, a net
flux of angular momentum is transport radially outward in the
disc. Conceptually, this process should cause the spinning SMBH
to gradually slow down, entirely analogous to the spin-down of
a rotating neutron star by emitting electromagnetic waves that
carry a net angular momentum away.

\section{Summary}

In this paper, we have explored and analyzed stationary perturbation
structures in coplanarly magnetized razor-thin scale-free discs in a
more general manner. To be specific, we start from a rotational and
magnetized background equilibrium of axisymmetry that is dynamically
self-consistent with a surface mass density $\Sigma_0\propto
r^{-\alpha}$, a rotation curve $v_0\propto r^{-\beta}$, a purely
azimuthal (ring) magnetic field $B_0\propto r^{-\gamma}$ with a
vertically integrated barotropic equation of state in the form of
$\Pi=K\Sigma^n$. The radial force balance at all radii in an MHD
disc implies several simple relationships among these power indices
$\alpha$, $\beta$, $\gamma$ and $n$, as stated explicitly in
equation (\ref{ParaRelation}). Without loss of generality, we can
simply use $\beta$ as an independent parameter to specify properties
of an important class of radial variations for rotational MHD
background equilibria. The allowed range of $\beta$ falls within the
interval $(-1/4,1/2)$ (Syer \& Tremaine 1996; Shen \& Lou 2004a, b)
with the special case of $\beta=0$ corresponding to a magnetized
singular isothermal disc (MSID) system (Shu et al. 2000; Lou 2002;
Lou \& Fan 2002; Lou \& Zou 2004a, b; Lou \& Wu 2004). For clarity
and convenience of our analysis, we introduced several dimensionless
parameters. The first parameter is the partial disc parameter
$F\equiv\phi/\phi_T$ for the ratio of the gravitational potential
($\phi$) arising from the disc to that ($\phi_T$) arising from the
entire system (Syer \& Tremaine 1996; Shu et al. 2000; Lou 2002)
including an axisymmetric halo mass distribution that is involved
in the background equilibrium but is presumed to be unresponsive
to coplanar MHD disc disturbances (a massive dark matter halo is
an example in mind); for a partial disc system, we have $0<{F}<1$,
while for a full disc system, we have ${F}=1$ by conventional
definitions. The second parameter is an effective rotational Mach
number $D$ defined as the ratio of the disc rotation speed to the
sound speed yet with an additional scaling factor
$(1+2\beta)^{-1/2}$; $D$ is constant at all radii. The third
parameter $q$ is a measure for the background ring magnetic field
strength defined as the ratio of the azimuthal Alfv\'en speed to
the sound speed; and $q$ remains constant at all radii. For
$1/4<\beta<1/2$, there is no constraint on $q$ and the net
Lorentz force of the background azimuthal magnetic field is
radially outward (the magnetic pressure force is stronger than the
magnetic tension force) to resist the self-gravitation; for
$\beta=1/4$, the background azimuthal magnetic field is force-free
(the magnetic pressure and tension forces cancel each other
exactly); for $-1/4<\beta<1/4$, the net Lorentz force of the
background azimuthal magnetic field points radially inward (the
magnetic pressure force is weaker than the magnetic tension force)
to aggravate the self-gravity and thus a restriction [i.e.,
inequality (\ref{restrt})] is imposed on the magnetic field
strength in order to guarantee that the surface mass density be
positive in equation (\ref{PropEquilib}). Inequality
(\ref{restrt}) provides a necessary physical criterion for
plausible $D^2$ solutions.

With such a rotational MHD background equilibrium chosen, we introduce
coplanar MHD disturbances to construct global stationary perturbation
structures as viewed in an inertial frame of reference. Perturbation
variables are expressed in terms of Fourier components for either
aligned or logarithmic spiral pattern forms (Syer \& Tremaine 1996;
Shu et al. 2000; Lou 2002; Lou \& Shen 2003; Shen \& Lou 2003; Lou
\& Zou 2004a). We derive analytical solutions for stationary
coplanar MHD perturbation configurations in both aligned and
logarithmic spiral cases. We now summarize key results below.


\begin{enumerate}
\item[(i)]{Cases of Aligned Coplanar MHD Perturbations}
\vskip 0.2cm

For aligned cases, we choose perturbation variables that bear the
same power-law dependence in radius $r$ as the background MHD
equilibrium does. For the special aligned case of $m=0$, the
stationary coplanar MHD perturbations actually describe
alternative equilibria to the axisymmetric background equilibrium
with proper rescaling factors (Shu et al. 2000; Lou 2002; Lou \&
Zou 2004a) and are therefore somewhat trivial.

For aligned cases of $m\ge 1$, we derive a quadratic equation
(\ref{aligned}) in terms of $y\equiv D^2$. The resulting positive
$D^2$ is the rotational parameter required for sustaining a
stationary nonaxisymmetric MHD perturbation configuration. Other
parameters involved in quadratic equation (\ref{aligned}) include
$\beta$, $m$, ${F}$ and $q$. For a general combination of
$\beta$, $m$, ${F}$ and $q$, the determinant $\Delta$ of
quadratic equation (\ref{aligned}) may not be always positive
definite. On the other hand, for specific values of $\beta$,
equation (\ref{aligned}) always gives two different real
solutions (see Appendix A). Once we prescribe parameters $\beta$,
$m$, ${F}$ and $q$, the two solutions of $D^2$ can be
readily determined. For physical solutions, $D^2$ must be
positive and satisfy inequality (\ref{restrt}) simultaneously.

As a specific case study, we examined the case of $\beta=1/4$.
Here the background azimuthal magnetic field remains force-free
and the $D^2$ solution is free from constraint (\ref{restrt}). It
has been further shown in Appendix A that for this $\beta=1/4$
case, equation (\ref{aligned}) always has two distinct real
solutions, the plus-signed solution $y_1$ and the minus-signed
solution $y_2$, as contained in equation (\ref{Solalign}). All
these aspects contribute to a significant simplification of the
relevant computational procedures. We derive two branches of
$D^2$ solutions and need to require only $D^2\ge 0$. There is
yet a qualitative
difference between the $m=1$ case and the $m\ge 2$ cases. For
$m=1$, there exists one critical value ${F}_c\simeq 0.6842$; for
${F}$ below ${F}_c$, $y_1$ and $y_2$ are negative and positive,
respectively, while ${F}$
above ${F}_c$, both branches of $D^2$ solutions are positive
and thus physical. For $m\ge 2$ in comparison, both branches of $D^2$
solutions remain always positive and thus physical. Moreover, $y_1$
remains always to be the upper branch for azimuthal stationary fast
MHD density waves and represents the hydrodynamic counterpart of
the unmagnetized single disc case; complementarily, $y_2$ remains
always to be the lower branch for azimuthal stationary slow MHD
density waves (Fan \& Lou 1996; Lou \& Fan 1998a) and is caused
by the very presence of the azimuthal magnetic field (detailed
computations reveal that $y_2$ must be smaller than $2/3$ which
seems to be too small for galactic applications). When a disc
rotates sufficiently fast, it could only support the upper $y_1$
stationary MHD configurations.
An increase of magnetic field parameter $q$ will raise both physical
branches of $D^2$ solutions as displayed in Fig. 1. The variation
trend of the upper $y_1$ branch appears to be more sensitive.
Therefore for a much stronger magnetic field, a sufficiently rapidly
rotating disc may sustain stationary coplanar MHD configurations in
terms of the upper $y_1=D^2$ branch.

\vskip 0.2cm
\item[(ii)]{Cases of Logarithmic Spiral
Configurations for Coplanar MHD Perturbations }
\vskip 0.2cm

For cases of stationary spiral perturbation structures, we choose
coplanar MHD perturbations in terms of Kalnajs logarithmic spirals
(Kalnajs 1971; Lemos et al. 1991; Shu et al. 2000; Lou 2002; Lou
\& Fan 2002; Lou \& Shen 2003; Lou \& Zou 2004a; Lou \& Wu 2004).
For the special $m=0$ case with radial propagations,
it happens that inequality (\ref{restrt}) is automatically
satisfied and the stationary coplanar MHD configurations with
$y\equiv D^2>0$ obtained here represent marginal stability curves
(Syer \& Tremaine 1996; Shu et al. 2000; Shen \& Lou 2003, 2004b;
Lou \& Zou 2004b). Only when the rotational
parameter $D^2$ falls within a specific finite range can the
magnetized disc be stable against axisymmetric disturbances at all
wavelengths. A disc with too slow a rotation speed will succumb to
Jeans' instability in the collapse regime corresponding to long
wavelength perturbations, while a disc with too fast a rotation
speed will suffer the ring-fragmentation instability corresponding
to relatively short wavelength perturbations (Safronov 1960;
Toomre 1964; Lemos et al. 1991; Lou \& Fan 1998a, 2000; Shu et al.
2000; Shen \& Lou 2003, 2004a; Lou \& Zou 2004b). In comparison
to a hydrodynamics disc system, the enhancement of azimuthal
magnetic field tends to suppress ring-fragmentation instabilities
as expected from the perspective of the MHD $Q$ parameter (Lou \&
Fan 1998a; Lou 2002; Lou \& Zou 2004a, b), while an enhancement of
azimuthal magnetic field
tends to suppress Jeans' collapse instabilities for
$-1/4<\beta<1/4$, to aggravate Jeans' collapse instabilities for
$1/4<\beta<1/2$ and to bear no effect on Jeans' collapse
instabilities when $\beta=1/4$. We have provided intuitive
interpretations for the roles of magnetic field earlier.

In parallel with the aligned cases, we derive quadratic equation
(\ref{spiral}) in terms of $y\equiv D^2$ for logarithmic spiral
cases of $m\ge 1$. In addition to the four parameters $\beta$,
$m$, ${F}$ and $q$, we now have one more parameter $\xi$ for the
radial wavenumber. While the determinant $\Delta$ of quadratic
equation (\ref{spiral}) may not be always positive in general, we
prove in Appendix A that $\Delta>0$ for the specific $\beta=1/4$
case. Once the five parameters $\beta$, $m$, ${F}$, $q$ and $\xi$
are specified, the $D^2$ solutions can be readily obtained. For
physical solutions, $D^2$ must be non-negative and also satisfy
inequality (\ref{restrt}).

As an example of illustration, we study the specific $\beta=1/4$
case inequality (\ref{restrt}) being satisfied automatically. Not
surprisingly, the aligned cases and the logarithmic spiral cases
parallel with each other very well, because of a mere change of
wave propagation direction. For the special spiral case of $m=1$,
there again exists the same critical ${F}_c\simeq 0.6842$
for ${F}$ value as in the aligned $m=1$ case; for ${F}
<{F}_c$, there is one critical point of $\xi_c$ where
$D^2\equiv y_1$ diverges. More specifically, we have unphysical
$y_1<0$ for $0<\xi<\xi_c$, while we have physical $y_1>0$ for
$\xi>\xi_c$. There is no divergent point for $D^2$ solution
branch and we always have $y_2\ge 0$ ($y_2\equiv 0$ when $q=0$).
The solution structures become simpler for $m\ge 2$ cases when
both $D^2$ branches of solutions are non-negative and there is
no divergent point of $D^2\equiv y_1$. Moreover, $y_1$ remains
always to be the upper branch for stationary spiral fast MHD
density waves and is the hydrodynamic counterpart for those in
the unmagnetized single disc case, while $y_2$ remains always
to be the lower branch for stationary spiral slow MHD density
waves (Fan \& Lou 1996) and is caused by the very presence of
the magnetic field (detailed computations reveal $y_2$ must be
smaller than $2/3$).

\vskip 0.2cm
\item[(iii)]{Phase Relationships among
Coplanar MHD Perturbation Variables}
\vskip 0.2cm

So far, we have constructed global stationary coplanar MHD
configurations for both aligned and logarithmic spiral cases in
magnetized discs with a range of rotation curves. As expected, the
inclusion of azimuthal magnetic field gives rise to one additional
distinct physical solution for stationary slow MHD density waves
to the problem (Fan \& Lou 1996; Lou \& Fan 1998a; Lou 2002; Lou
\& Zou 2004a); the Alfv\'enic modes are excluded by our
restricted consideration for coplanar MHD perturbations. The
two $D^2$ branches of solutions are at least
mathematically reasonable and may both have applications to
astrophysical disc systems such as magnetized disc galaxies and so
forth. These devised MHD disc problems are highly idealized with
yet precious global analytical perturbation solutions, and
conceptually still belong to a subclass of more general
time-dependent MHD perturbation solutions. For applications to
magnetized spiral galaxies, it would be of considerable interest
to discuss pertinent physical aspects and identify relevant
observational diagnostics.

For physical solutions of $D^2>0$, the $y_1$ branch would be
applicable to MHD discs with relatively fast rotation while the
$y_2$ branch would be applicable to MHD discs with relatively slow
rotation in general. For spatial phase relationships, the two
branches of $D^2>0$ solutions give different results. We take the
specific $\beta=1/4$ case as an example of illustration. For
corresponding aligned cases, the plus-solution $y_1$ will lead
to $\hbox{i}R/S>0$ and the minus-solution $y_2$ will lead to
$\hbox{i}R/S<0$; there are no azimuthal magnetic field
perturbations for aligned $\beta=1/4$ cases. For tightly wound
logarithmic spiral cases, perturbations of both radial and
azimuthal magnetic field are approximately in phase with the
perturbation enhancement of the surface mass density for the $y_1$
branch, while they are approximately out of phase for the $y_2$
branch. For open logarithmic spirals, the radial and azimuthal
magnetic field perturbations are either ahead of or lagging behind
the enhancement of the surface mass by a significant phase
difference. These phase relationships together with those
associated with flow perturbations (Visser 1980a, b) are valuable
to interpret the spatial phase shifts between optical arms and
magnetic arms in lopsided, barred or other spiral galaxies
(Mathewson et al. 1972; Neininger 1992; Beck \& Hoernes 1996; Beck
et al. 1996; Fan \& Lou 1996; Lou \& Fan 1998, 2000, 2002; Frick
et al. 2000, 2001; Lou et al. 2002).

\vskip 0.2cm
\item[(iv)]{Angular Momentum Transfer in Steady Logarithmic
Spirals for Coplanar MHD Perturbation Configurations}
\vskip 0.2cm

Stationary logarithmic spiral configurations of MHD density waves
carry constant angular momentum flux either outward or inward
associated with the advective transport, the gravity torque and
the magnetic torque (Lynden-Bell \& Kalnajs 1972; Goldreich \&
Tremaine 1978; Fan \& Lou 1999). Therefore, the net angular
momentum flux in the entire disc system is conserved and there
must be either a source or a sink at the disc center. For the
$y_1$ branch with $D^2>0$, we find that the total angular momentum
flux is inward for leading (i.e., $\xi>0$) spiral MHD density
waves and is outward for trailing (i.e., $\xi<0$) spiral MHD
density waves. For the $y_2$ branch in comparison, the total
angular momentum flux is outward for leading (i.e., $\xi>0$)
spiral MHD density waves and is inward for trailing (i.e.,
$\xi<0$) spiral MHD density waves. As a real MHD disc system
typically rotates at a relatively large $D^2$ and inequality
(\ref{restrt}) can also rule out small $D^2$ for $\beta\neq1/4$,
the lower $y_2$ branch might be rarely useful in galactic
applications. In either case, however, the net mass accretion
rate is zero since if we focus on materials between two
concentric circles, at any time the influent angular momentum
from the inner ring is equal to the effluent angular momentum
from the outer ring such that materials inside this belt neither
gain nor lose any angular momentum and therefore no net mass
accretion occurs.

\end{enumerate}

\section*{Acknowledgments}
This research has been supported in part by the ASCI Center for
Astrophysical Thermonuclear Flashes at the University of Chicago
under Department of Energy contract B341495, by the Special Funds
for Major State Basic Science Research Projects of China, by the
Tsinghua Center for Astrophysics, by the Collaborative Research
Fund from the National Natural Science Foundation of China (NSFC)
for Young Outstanding Overseas Chinese Scholars (NSFC 10028306) at
the National Astronomical Observatory, Chinese Academy of Sciences,
by NSFC grant 10373009 at the Tsinghua University, and by the
Yangtze Endowment from the Ministry of Education through the
Tsinghua University. Affiliated institutions of Y.Q.L. share this
contribution.

\appendix
\section{Basic Properties of Determinant $\Delta$}

For $m\ge 1$, we first examine the solution criterion of quadratic
equation (\ref{aligned}) for the aligned case. The determinant
$\Delta$ of quadratic equation (\ref{aligned}) is
\begin{equation}\label{A01}
\begin{split}
\Delta&\equiv C_1^2-4C_2C_0\\
&=c_2q^4+c_1q^2+c_0\ ,
\end{split}
\end{equation}
where the three coefficients $c_j$ ($j=0,1,2$)
are defined explicitly by
\begin{equation}\label{A02}
\begin{split}
c_2\equiv\  &\bigg[\frac{(3-4\beta)m^2-(1-4\beta)(1-6\beta+4\beta^2)}
{2(1+2\beta)}{F}{\cal C}{\cal P}_m+m^2
-(1-4\beta)(2-\beta)\bigg]^2\\
&\qquad\qquad\qquad\qquad\qquad
+\frac{(1-4\beta)}{(1+2\beta)}
\bigg[\frac{-2m^2+(1-4\beta)(1-2\beta)}{(1+2\beta)}{F}
{\cal C}{\cal P}_m+1-4\beta\bigg]{\cal H}_m\ ,\\
c_1\equiv\  &2(1-{F}{\cal C}{\cal P}_m)\bigg\{{\cal A}_m
\bigg[\frac{(3-4\beta)m^2-(1-4\beta)(1-6\beta+4\beta^2)}{2(1+2\beta)}
{F}{\cal C}{\cal P}_m+m^2-(1-4\beta)(2-\beta)\bigg]\\
&\qquad\qquad\qquad\qquad\qquad
+\frac{-2m^2+(1-4\beta)(1-2\beta)}{(1+2\beta)}{\cal H}_m\bigg\}\ ,\\
c_0\equiv\  &{\cal A}_m^2(1-{F}{\cal C}{\cal P}_m)^2\ ,
\end{split}
\end{equation}
respectively. For the quadratic expression (\ref{A01}) in terms of
$q^2$, we further introduce another determinant $\Delta_1$ such that
\begin{equation}
\Delta_1\equiv c_1^2-4c_2c_0=-\frac{4{\cal H}_m(1-{F}
{\cal C}{\cal P}_m)^2}{(1+2\beta)}[3m^2-2(1-\beta)(1-4\beta)]
[(1+8\beta)m^2-(1-4\beta)(1+4\beta-8\beta^2)]\ .
\end{equation}

The determinant $\Delta$ of quadratic equation (\ref{aligned})
in terms of $D^2$ can be either positive or negative in general
for different parameter combinations of $\beta$, $m$, $q$ and
${F}$. It is fairly straightforward to obtain $D^2$ solutions
by assigning values of all relevant parameters. In particular,
for one important case of $\beta=0$ (i.e., the MSID case with
a flat rotation curve and an isothermal equation of state), the
condition that $\Delta\ge 0$ for $m\ge 1$ and $\Delta=0$ when
$m=1$ holds for all values of $q^2$ in a full MSID with ${F}=1$
and the condition that $\Delta>0$ for $m\ge 1$ holds in a
partial MSID with $0<{F}<1$ (Lou 2002).

For the magnetic force-free case of $\beta=1/4$ with a declining
rotation curve with increasing $r$ and a polytropic index $n=4/3$,
we have $\Delta>0$ for $m\ge 1$ and an arbitrary $q>0$ in both full
and partial magnetized discs. We now proceed to prove this statement
rigorously. For $\beta=1/4$ of the aligned case, we explicitly have
\begin{equation}
c_2\equiv [m^2(2{F}{\cal C}{\cal P}_m/3+1)]^2\ ,
\qquad
c_1\equiv 2(1-{F}{\cal C}{\cal P}_m)m^2
[{\cal A}_m(2{F}{\cal C}{\cal P}_m/3+1)-4{\cal H}_m/3]\ ,
\qquad
c_0\equiv {\cal A}_m^2(1-{F}{\cal C}{\cal P}_m)^2\ ,
\end{equation}
where $0<1-{F}{\cal C}{\cal P}_m<1$ and ${\cal A}_m>0$ for
$\beta=1/4$ and $\Delta_1$ introduced above simply reduces to
\begin{equation}
\Delta_1=-24{\cal H}_m(1-{F}{\cal C}{\cal P}_m)^2m^4\ .
\end{equation}
It is apparent that for ${\cal H}_m>0$, we have $\Delta_1<0$. As
$c_2>0$, it follows immediately that $\Delta>0$ in definition
(\ref{A01}). On the other hand for ${\cal H}_m\le 0$, we have
$c_2>0,\ c_1>0,\ c_0>0$ and therefore $\Delta>0$ by definition
(\ref{A01}).

For $\beta=1/4$ in the logarithmic spiral case, we again have
the determinant $\Delta>0$ for quadratic equation (\ref{spiral}),
where in parallel with definitions (\ref{A01}) and (\ref{A02}) of
the aligned case, the relevant parameters in the spiral case are
\begin{equation}
c_2=\bigg[\bigg(m^2+\xi^2\bigg)\bigg(\frac{2}{3}{F}
{\cal C}{\cal N}_m+1\bigg)\bigg]^2\ ,\ c_1=2(1-{F}{\cal C}
{\cal N}_m)(m^2+\xi^2)\bigg[{\cal A}_m\bigg(\frac{2}{3}{F}
{\cal C}{\cal N}_m+1\bigg)-\frac{4}{3}{\cal H}_m\bigg]\ ,\
c_0={\cal A}_m^2(1-{F}{\cal C}{\cal N}_m)^2\ ,
\end{equation}
with $0<1-{F}{\cal C}{\cal N}_m<1$ and ${\cal A}_m>0$ for
$\beta=1/4$. And $\Delta_1$ simply reduces to
\begin{equation}
\Delta_1=-\frac{32}{9}{\cal H}_m(1-{F}{\cal C}{\cal
N}_m)^2(m^2+\xi^2)^2(3\xi^2+27/4)\ .
\end{equation}
Hence, as the aligned $\beta=1/4$ case, it is apparent that for
${\cal H}_m>0$, we have $\Delta_1<0$. As $c_2>0$, it follows
immediately that $\Delta>0$. On the other hand for ${\cal H}_m
\le 0$, we have $c_2>0,\ c_1>0,\ c_0>0$ and therefore $\Delta>0$ .


\section{ Phase Relationships among Perturbation Variables }

In this appendix, we first provide a proof that for the aligned
case of $\beta=1/4$, the physical $D^2$ solutions of the plus
$y_1>0$ and the minus $y_2>0$ determined from quadratic
equation (\ref{q2aligned}) correspond to the factor
$D^2-(1-{F}{\cal C}{\cal P}_m)/({F}{\cal C} {\cal P}_m+3/2)$
being positive and negative,
respectively. According to Appendix A, the determinant
$\Delta$ for both aligned and spiral cases remains
always positive for the case of $\beta=1/4$.

We start from equation (\ref{q2aligned}) by
casting it in the following compact form of
\begin{equation}
\wp_3=\frac{y(y-\wp_1)}{y-\wp_2}\ ,
\end{equation}
where $y\equiv D^2$ and in reference to equation (\ref{q2aligned}),
$\wp_1$, $\wp_2$ and $\wp_3$ are constants defined explicitly by
\begin{equation}
\wp_1\equiv\frac{(1-{F}{\cal C}{\cal P}_m)}
{{F}{\cal C}{\cal P}_m
+3(m^2-3/2)/(2m^2+3/2) }\ ,\quad
\wp_2\equiv\frac{(1-{F}{\cal C}{\cal P}_m)}
{({F}{\cal C}{\cal P}_m+3/2)}\ ,\quad
\wp_3\equiv\frac{m^2q^2(2{F}{\cal C}{\cal P}_m/3+1)}
{(m^2+3/4)[{F}{\cal C}{\cal P}_m+3(m^2-3/2)
/(2m^2+3/2)] }\ .
\end{equation}
It is apparent that $\wp_2$ is always positive. While $\wp_1$ and
$\wp_3$ remain always positive for $m\ge 2$, their signs depend on
the value of ${F}$ for the special case of $m=1$. In fact,
both $\wp_1$ and $\wp_3$ diverge at a critical value of ${F}$
for the $m=1$ case as already discussed in subsection 3.1.3.

Let us first analyze the simpler cases of $m\ge 2$. As
$\wp_1>0\ ,\ \wp_2>0\ ,\ \wp_3>0$ and $\wp_1>\wp_2$,
we have two $D^2$ solutions
\begin{equation}\label{B2}
\begin{split}
y_1=\frac{\wp_1+\wp_3+[(\wp_1+\wp_3)^2-4\wp_2\wp_3]^{1/2}}{2}&>
\frac{\wp_2+\wp_3+|\wp_2-\wp_3|}{2}\ge \wp_2\ ;\\
y_2=\frac{\wp_1+\wp_3-[(\wp_1+\wp_3)^2-4\wp_2\wp_3]^{1/2}}{2}
&=\frac{\wp_1+\wp_3-[(\wp_1+\wp_3-2\wp_2)^2+4\wp_2(\wp_1-\wp_2)]^{1/2}}{2}\\
&<\frac{\wp_1+\wp_3-|\wp_1+\wp_3-2\wp_2|}{2}\le \wp_2\ .
\end{split}
\end{equation}

We next examine the $m=1$ case. The situation becomes somewhat
involved since $\wp_1<0,\ \wp_3<0$, $|\wp_1|>\wp_2>0$ when
$0<{F}<{F}_c$ where the critical ${F}$ value
${F}_c\equiv 3/[7{\cal C}{\cal P}_1(1/4)]\simeq 0.6842$
is determined from the condition ${\cal H}_1=0$ (see subsection
3.1.3). On the other hand, we have $\wp_1>\wp_2>0,\ \wp_3>0$
when ${F}>{F}_c$ in general. For the aligned
case\footnote{Note these expressions of $y_1$ and $y_2$ are
consistent with the general solution form (\ref{Solalign})
for the aligned case.} and for $0<{F}<{F}_c$, we
therefore have two $D^2$ solutions satisfying inequalities
\begin{equation}
\begin{split}
y_1&=\frac{\wp_1+\wp_3
-[(\wp_1+\wp_3)^2-4\wp_2\wp_3]^{1/2}}{2}<0\ ;\\
0<y_2&=\frac{\wp_1+\wp_3+[(\wp_1+\wp_3)^2-4\wp_2\wp_3]^{1/2}}{2}
\\
&=\frac{\wp_1+\wp_3+[(\wp_1+\wp_3-2\wp_2)^2
 +4\wp_2(\wp_1-\wp_2)]^{1/2}}{2}\\
&<\frac{\wp_1+\wp_3+|\wp_1+\wp_3-2\wp_2|}{2}=\wp_2\ ,
\end{split}
\end{equation}
and for ${F}>{F}_c$ the situation
remains the same as (\ref{B2}).

For the spiral case of $\beta=1/4$ in parallel, we can
repeat this same procedure of analysis. As before, we
cast quadratic equation (\ref{beta0d25s}) in the form of
\begin{equation}\label{spiral0d25}
\begin{split}
&\qquad
\frac{(m^2+\xi^2)q^2(2{F}{\cal C}{\cal N}_m/3+1)}
{(m^2+\xi^2+3/4)[{F}{\cal C}{\cal N}_m
+3(m^2-3/2)/(2m^2+2\xi^2+3/2)]} \\
&\qquad\qquad\qquad\qquad\qquad\qquad\qquad\qquad
=\frac{D^2\{ D^2-(1-{F}{\cal C}{\cal N}_m)/
[{F}{\cal C}{\cal N}_m+3(m^2-3/2)/(2m^2+2\xi^2+3/2)]\} }
{D^2-(1-{F}{\cal C}{\cal N}_m)/
({F}{\cal C}{\cal N}_m+3/2)}\ ,
\end{split}
\end{equation}
which yields two distinct $D^2$ solutions according to
the analysis in subsection 3.2.3. The following treatment
is identical to that for the aligned $\beta=1/4$ case
described in the first part of this appendix.

\section{Angular Momentum Flux}

We here provide a proof to show that the angular momentum
flux associated with stationary coplanar MHD perturbations due
to the gravity torque $\Lambda^{G}$, the advective transport
$\Lambda^{A}$ and the magnetic torque $\Lambda^{B}$ remains
constant at all radii for the entire range $-1/4<\beta<1/2$.

We start from expression (\ref{rmphi}) for $\Re(\phi_1)$ to
derive the following two expressions in a straightforward
manner:
\begin{equation}
\begin{split}
&\frac{\partial\Re(\phi_1)}{\partial \theta}=2\pi
Gm\sigma\int_0^{\infty}dke^{-k|z|}J_m(kr)
\int_0^{\infty}r_a^{-1/2}\sin(m\theta-\xi\ln r_a)J_m(kr_a)dr_a\
,\\
&\frac{\partial\Re(\phi_1)}{\partial r}=-2\pi
G\sigma\int_0^{\infty}dk^{\prime}e^{-k^{\prime}|z|}
\bigg[-k^{\prime}J_{m+1}(k^{\prime}r)
+\frac{m}{r}J_m(k^{\prime}r)\bigg]J_m(k^{\prime}r)
\int_0^{\infty}r_a^{\prime\ -1/2}\cos(m\theta-\xi\ln
r_a^{\prime})J_m(k^{\prime}r_a^{\prime})dr_a^{\prime}\ .
\end{split}
\end{equation}
Then in integral (\ref{gravangm}), we first
integrate the kernel over $\theta$ to yield
\begin{equation}
-4\pi^3G^2\sigma^2mr\int_0^{\infty}\int_0^{\infty}F(k,k^{\prime})
e^{-(k+k^{\prime})|z|}J_m(kr)J_m(k^{\prime}r)
\bigg[-k^{\prime}J_{m+1}(k^{\prime}r)
+\frac{m}{r}J_m(k^{\prime}r)\bigg]dkdk^{\prime}\ ,
\end{equation}
where
\begin{equation}
F(k,k^{\prime})\equiv\int_0^{\infty}\int_0^{\infty}
r_a^{-1/2}r_a^{\prime\ -1/2}J_m(kr_a)J_m(k^{\prime}r_a^{\prime})
\sin\bigg[\xi\ln\bigg(\frac{r_a^{\prime}}
{r_a}\bigg)\bigg]dr_adr_a^{\prime}\ .
\end{equation}
We next integrate over the vertical
coordinate $z$ to yield the final result
\begin{equation}\label{final}
-4\pi^3G^2\sigma^2mr\int_0^{\infty}
\int_0^{\infty}\frac{F(k,k^{\prime})}{(k+k^{\prime})}
J_m(kr)J_m(k^{\prime}r)\bigg[-k^{\prime}J_{m+1}(k^{\prime}r)
+\frac{m}{r}J_m(k^{\prime}r)\bigg]dkdk^{\prime}\ .
\end{equation}
By introducing the following integral transformation
\begin{equation}
kr\rightarrow x,\ \qquad k^{\prime}r\rightarrow x^{\prime},\ \qquad
r_a\rightarrow rr_b,\ \qquad r_a^{\prime}\rightarrow rr_b^{\prime}\ ,
\end{equation}
we reduce expression (\ref{final}) to
\begin{equation}
-4\pi^3G^2\sigma^2m\int_0^{\infty}
\int_0^{\infty}\frac{M(x,x^{\prime})}{x+x^{\prime}}
J_m(x)J_m(x^{\prime})\bigg[-x^{\prime}J_{m+1}(x^{\prime})
+mJ_m(x^{\prime})\bigg]dxdx^{\prime}\ ,
\end{equation}
where
\begin{equation}
M(x,x^{\prime})\equiv\int_0^{\infty}\int_0^{\infty}r_b^{-1/2}
r_b^{\prime\ -1/2}J_m(xr_b)J_m(x^{\prime}r_b^{\prime})
\sin\bigg[\xi\ln\bigg(\frac{r_b^{\prime}}{r_b}\bigg)
\bigg]dr_bdr_b^{\prime}={\cal N}_mx^{-1/2}x^{\prime\
-1/2}\sin\bigg[\xi\ln\bigg(\frac{x}{x^{\prime}}\bigg)\bigg]\ ,
\end{equation}
where ${\cal N}_m$ is the Kalnajs function (Kalnajs 1971).

With these technical preparations, we finally determine the
angular momentum flux associated with coplanar stationary
MHD perturbations due to the gravity torque as
\begin{equation}
\Lambda^{G}=-m\pi^2G\sigma^2\Xi^G\ ,
\end{equation}
where
\begin{equation}\nonumber
\Xi^G\equiv\int_0^{\infty}\int_0^{\infty}\frac{M(x,x^{\prime})}
{(x+x^{\prime})}
J_m(x)J_m(x^{\prime})[-x^{\prime}J_{m+1}(x^{\prime})
+mJ_m(x^{\prime})]dxdx^{\prime}\ ,
\end{equation}
which is clearly independent of radius $r$. In the
tight-winding regime or the WKBJ limit (i.e., $\xi\gg1$),
we may directly apply the result of Fan \& Lou (1999,
see their eqn. 4.13) by noting $kr\equiv\xi$ such that
\begin{equation}
\Lambda^G=-\hbox{sgn}(\xi)m\pi^2G\sigma^2|\Theta_S|^2\ ,
\end{equation}
where $\Theta_S$ has been defined by equation (\ref{ThetaS});
here, $\Lambda^G$ is inward for leading ($\xi>0$) spiral waves
and outward for trailing ($\xi<0$) spiral waves. In general,
$\Xi^G$ can be calculated via numerical integrations and the
specific result is displayed in Fig. C1. For fairly open MHD
density waves ($\xi\simeq 1$) with $\Xi^G>0$, the angular
momentum flux associated with the gravity torque is therefore
inward for leading ($\xi>0$) spiral waves and outward for
trailing ($\xi<0$) spiral waves; with increasing $\xi\gsim 1$,
$\Xi^G$ falls off rapidly and oscillates around zero with the
envelope profile closely fitting the WKBJ result better and better.

\begin{figure}
\centering
\includegraphics[scale=0.42]{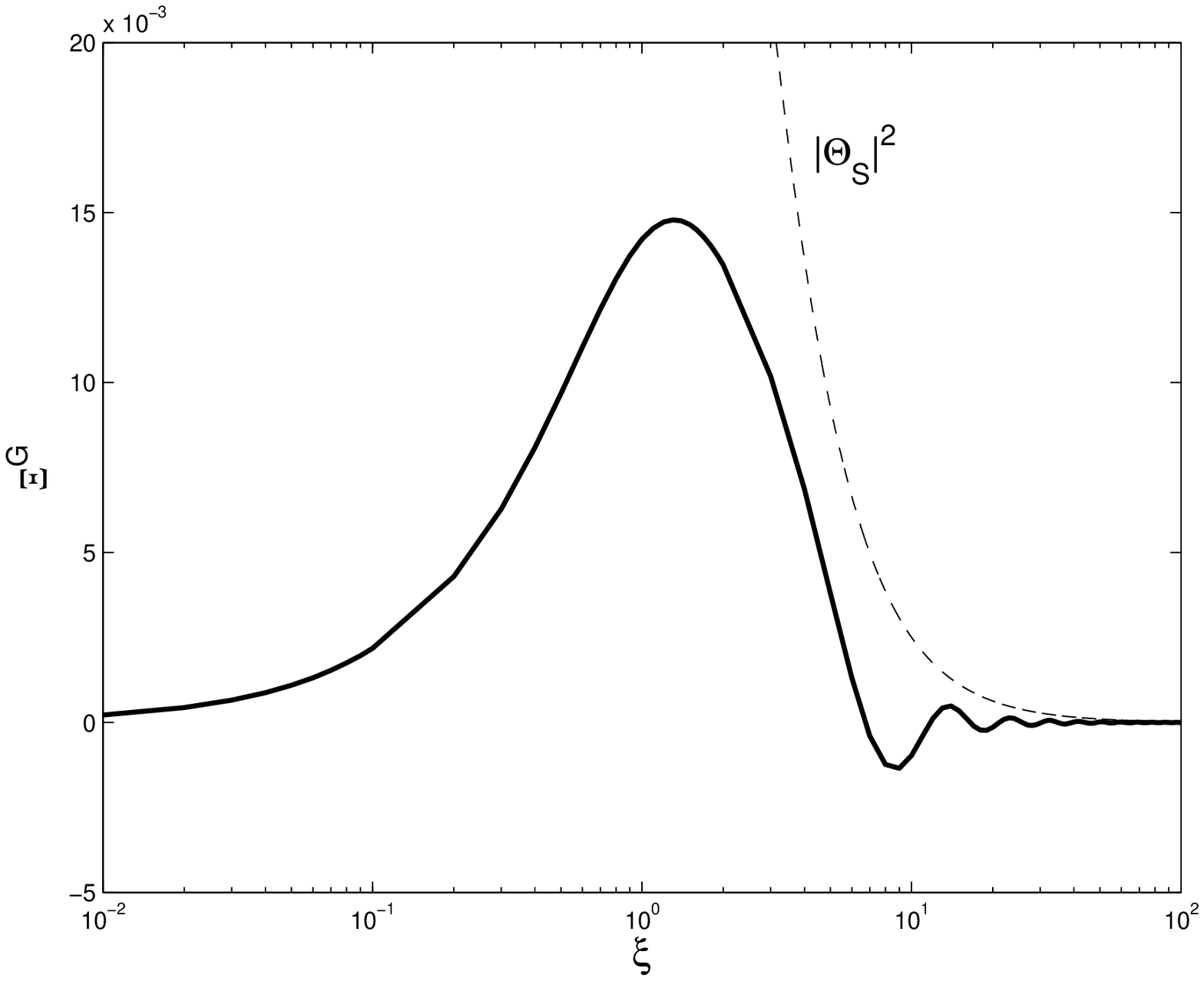}
\caption{Variation of $\Xi^G$ versus $\xi$, shown in heavy solid
line. We plot $|\Theta_S|^2$ in light dashed line for comparison.
For increasing $\xi\gsim 1$, $\Xi^G$ oscillates around zero and
its envelope profile fits the $|\Theta_S|^2$ curve with an
increasing accuracy.}
\end{figure}

\begin{figure}
\centering \subfigure[Variation of $\Xi^A$ versus $\xi$]{
\includegraphics[scale=0.42]{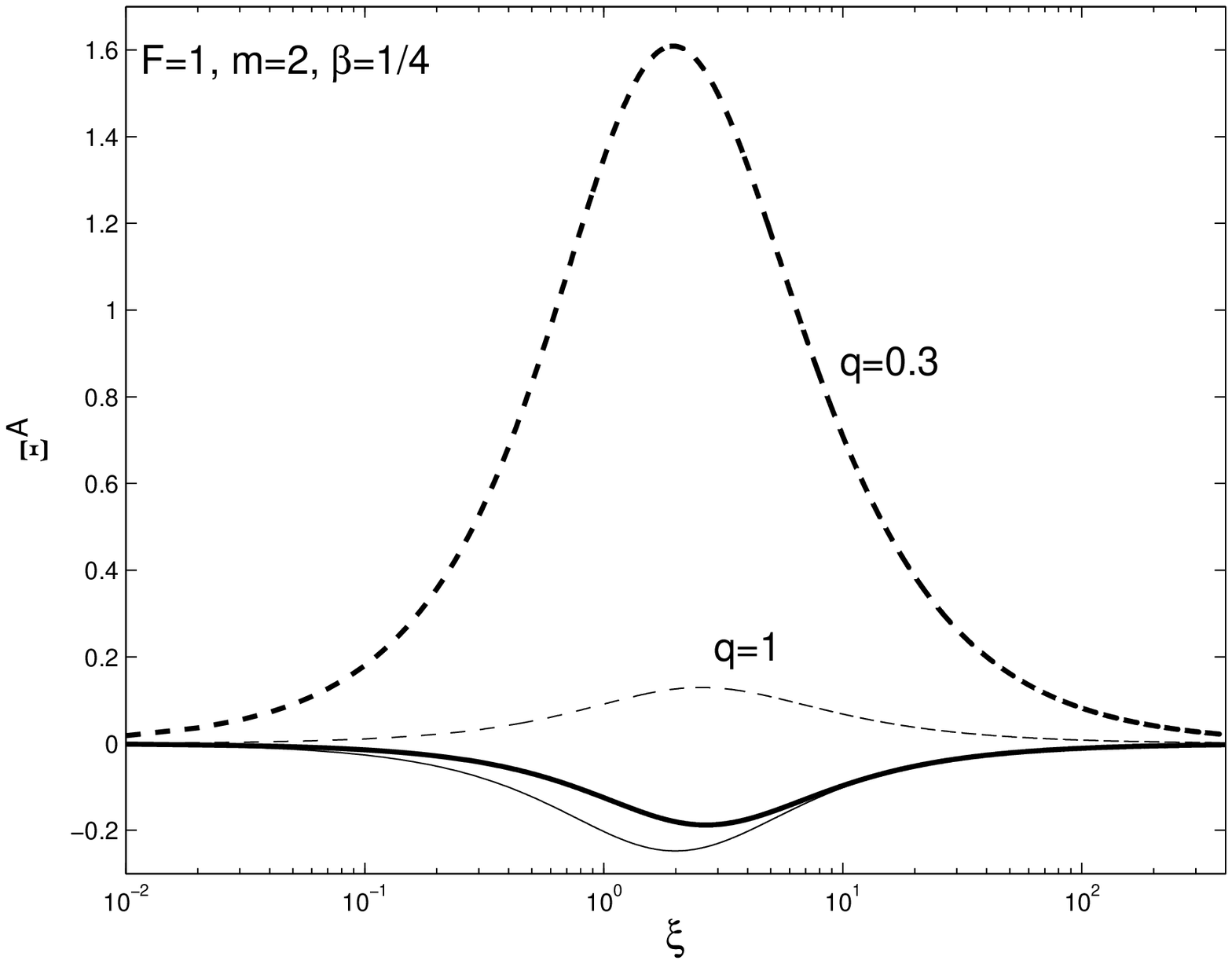}}
\subfigure[Variation of $\Xi^B$ versus $\xi$]{
\includegraphics[scale=0.42]{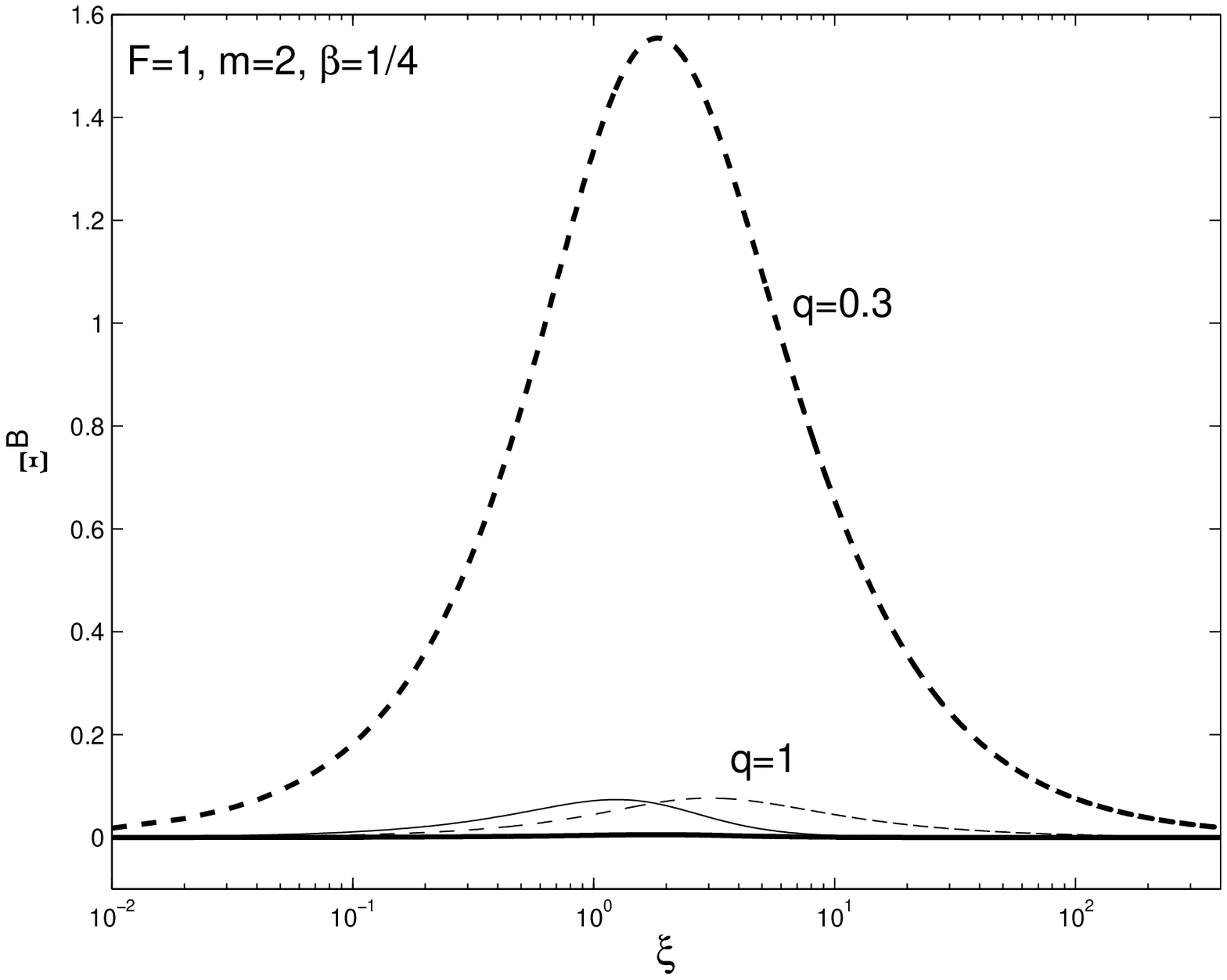}}
\caption{Variations of $\Xi^A$ and $\Xi^B$ versus $\xi$ for
the two branches of $y=D^2$ solution with $m=2$, $\beta=1/4$,
${F}=1$, and $q=0.3,\ 1$. Solid and dashed curves stand
for $y_1$ and $y_2$ branches, respectively, with heavy and light
curves for $q=0.3$ and $q=1$, respectively. Note that $\Xi^A$ is
positive and negative for $y_2$ and $y_1$ branches, respectively,
while $\Xi^B$ is positive for both $y=D^2$ branches.}
\end{figure}

Meanwhile, we compute the other two perturbation contributions
to the angular momentum flux, namely, those from the advective
transfer and the magnetic torque. For the part of advective
transport, we start from equation (\ref{adveangm}) and make use
of phase relationship (\ref{RelSURJZspiral}) to explicitly derive
\begin{equation}
\begin{split}
&\Re(u_1)=\frac{m\Omega\sigma\Re(\Theta_S)}
{\Sigma_0}r^{-1/2}\sin(\xi\ln r-m\theta)
+\frac{m\Omega\sigma\Im(\Theta_S)}
{\Sigma_0}r^{-1/2}\cos(\xi\ln r-m\theta)\ ,\\
&\Re(j_1/r)=-(\Omega\sigma /\Sigma_0)
[1-(1/2+\beta)\Re(\Theta_S)
-\xi\Im(\Theta_S)]r^{-1/2}\cos(\xi\ln r-m\theta)\\
&\qquad\qquad\qquad\qquad\qquad\qquad\qquad\
-(\Omega\sigma/\Sigma_0)[(1/2+\beta)\Im(\Theta_S)
-\xi\Re(\Theta_S)]r^{-1/2}\sin(\xi\ln r-m\theta)\ .
\end{split}
\end{equation}
It immediately follows that the advective torque is given by
\begin{equation}
\Lambda^{A}=r^2\Sigma_0\int_0^{2\pi}d\theta
\Re(u_1)\Re(j_1/r)=\frac{m\pi\sigma^2\Omega^2r}{\Sigma_0}
[\xi|\Theta_S|^2-\Im(\Theta_S)]=m\pi^2G\sigma^2\Xi^A\ ,
\end{equation}
where
\begin{equation}\nonumber
\Xi^A\equiv\frac{2D^2(1+2\beta)[\xi|\Theta_S|^2-\Im(\Theta_S)]}
{{F}{\cal C}[1+D^2-(1-4\beta)q^2/(2+4\beta)]}\ ,
\end{equation}
that is independent of radius $r$. By specifying parameters
$\beta$, $m$, ${F}$, $q$ and $\xi$, one finds that for
the $y_1$ branch $\xi|\Theta_S|^2-\Im(\Theta_S)<0$ for $\xi>0$
and $\xi|\Theta_S|^2-\Im(\Theta_S)>0$ for $\xi<0$.
Therefore, the advective angular momentum flux is radially inward
for leading stationary spiral density waves and radially outward
for trailing stationary spiral density waves for the $y_1$ branch
of $D^2$ solutions. For the $y_2$ branch in comparison, the
situation switches and the advective angular momentum flux is
outward for leading stationary spiral density waves and inward for
trailing stationary spiral density waves. Practically, to achieve
the $y_2$ branch of stationary logarithmic spiral configurations
requires a much lower $D^2$ which would be unusual for real
magnetized disc systems. An example of illustration for $m=2$,
${F}=1$ and $\beta=1/4$ is shown in panel (a) of Fig. C2.

For the momentum flux associated with the magnetic torque, we
start from equation (\ref{magnangm}) and make use of phase
relationship (\ref{RelSURJZspiral}) to first derive
\begin{equation}
\begin{split}
&\Re(b_r)=[mB_0\sigma\Re(\Theta_S)/\Sigma_0]
\ r^{-3/2}\sin(\xi\ln r-m\theta)
+[mB_0\sigma\Im(\Theta_S)/\Sigma_0]\ r^{-3/2}
\cos(\xi\ln r-m\theta)\ ,\\
&\Re(b_\theta)=[\xi B_0\sigma\Re(\Theta_S)
/\Sigma_0]\ r^{-3/2}\sin(\xi\ln r-m\theta)
+[\xi B_0\sigma\Im(\Theta_S)/\Sigma_0]
\ r^{-3/2}\cos(\xi\ln r-m\theta)\ .
\end{split}
\end{equation}
It immediately follows that the momentum flux $\Lambda^{B}$
associated with the magnetic torque is given by
\begin{equation}
\Lambda^{B}=-\frac{r^2}{4\pi}\int_0^{2\pi}
d\theta\int_{-\infty}^{\infty}dz\Re(b_r)\Re(b_\theta)
=-\frac{m\pi\xi\sigma^2C_A^2|\Theta_S|^2}{\Sigma_0r}
=-m\pi^2G\sigma^2\Xi^B\ ,
\end{equation}
where
\begin{equation}\nonumber
\Xi^B\equiv\frac{2q^2\xi |\Theta_S|^2}{{F}
{\cal C}[1+D^2-(1-4\beta)q^2/(2+4\beta)] }\ ,
\end{equation}
which is again independent of radius $r$; here, $\Lambda^{B}$
is inward for leading spiral density waves (i.e., $\xi>0$) and
outward for trailing spiral density waves (i.e., $\xi<0$). An
example of illustration for $m=2$, ${F}=1$ and $\beta=1/4$ is
shown in panel (b) of Fig. C2.

By comparing the three contributions to the total angular momentum
flux transport, we reach the following conclusions. First, the
contribution from the advective transport part tends to be more
dominant over those from the magnetic torque and the gravity torque.
Secondly, for the $y_1$ branch for stationary fast MHD density waves
that has a counterpart in the unmagnetized case, leading spiral
density waves ($\xi>0$) transport angular momentum inward and
trailing spiral density waves ($\xi<0$) transport angular momentum
outward. Thirdly, for the $y_2$ branch for stationary slow MHD
density waves caused by the very presence of the azimuthal magnetic
field, leading spiral density waves ($\xi>0$) carry angular
momentum outward and trailing spiral density waves ($\xi<0$)
carry angular momentum inward.
Note that all these results are only valid for stationary logarithmic
spiral MHD density waves. There might be some reasons to rule out
the $y_2$ branch in galactic contexts. For example, the $y_2$ branch
is of lower $D^2$ which cannot be sustained in real galactic discs.
Furthermore, for rotation curve parameter $\beta\neq1/4$, the $y_2$
branch may be ruled out by inequality (\ref{restrt}).

\end{document}